\documentclass[aps,prb,twocolumn,shortbibliography,superscriptaddress,article]{revtex4-1}
\usepackage{epsfig}
\usepackage{epstopdf}
\usepackage{amsmath}
\usepackage{amsfonts}
\usepackage{amssymb}
\usepackage{hyperref}
\usepackage{bm}
\usepackage{makecell}
\usepackage{rotating}
\usepackage{hyperref}
\usepackage{multirow}
\usepackage{graphicx}
\usepackage{booktabs}
\usepackage{siunitx}
\usepackage{makecell}

\usepackage{graphicx}
\usepackage{dcolumn}
\usepackage{bm}
\usepackage{color}

\usepackage{tikz,xcolor,hyperref}

\definecolor{lime}{HTML}{A6CE39}
\DeclareRobustCommand{\orcidicon}{%
	\begin{tikzpicture}
	\draw[lime, fill=lime] (0,0)
	circle [radius=0.16]
	node[white] {{\fontfamily{qag}\selectfont \tiny ID}};
	\draw[white, fill=white] (-0.0625,0.095)
	circle [radius=0.007];
	\end{tikzpicture}
	\hspace{-2mm}
}

\foreach \x in {A, ..., Z}{%
	\expandafter\xdef\csname orcid\x\endcsname{\noexpand\href{https://orcid.org/\csname orcidauthor\x\endcsname}{\noexpand\orcidicon}}
}

\begin{document}

\title{Tunability of the magnetic properties in Ni intercalated \\
transition metal dichalcogenide NbSe$_2$}

\author{Xujia Gong\orcidA}
\email{xgong@magtop.ifpan.edu.pl}
\affiliation{International Research Centre Magtop, Institute of Physics, Polish Academy of Sciences,
Aleja Lotnik\'ow 32/46, PL-02668 Warsaw, Poland}

\author{Amar Fakhredine\orcidF}
\affiliation{Institute of Physics, Polish Academy of Sciences, Aleja Lotnik\'ow 32/46, 02668 Warsaw, Poland}

\author{Carmine Autieri\orcidB}
\email{autieri@magtop.ifpan.edu.pl}
\affiliation{International Research Centre Magtop, Institute of Physics, Polish Academy of Sciences,
Aleja Lotnik\'ow 32/46, PL-02668 Warsaw, Poland}
\affiliation{SPIN-CNR, UOS Salerno, IT-84084 Fisciano (SA), Italy}

\date{\today}
\begin{abstract}
We study the magnetic and electronic properties of Ni-intercalated NbSe$_2$.
We calculate the magnetic exchanges of Ni$_x$NbSe$_2$ ($x = 1/3, 1/4,$ and $1$)
and find that the out-of-plane magnetic coupling depends on the Ni connectivity:
it is ferromagnetic when Ni atoms stack on top of each other, and antiferromagnetic otherwise.
Focusing on Ni$_{0.25}$NbSe$_2$, we identify a ground-state transition from a stripe antiferromagnetic
phase with Kramers degeneracy to a ferromagnetic phase above a critical Coulomb interaction U$_C$.
Spin--orbit coupling lowers U$_C$, aligns the easy axis along $z$, and stabilizes collinear AFM and FM states
over the competing 120$^\circ$ phase.
Ni intercalation also strongly modifies the electronic structure, replacing the $\Gamma$-point hole pocket
of pristine NbSe$_2$ with an electron pocket and shifting the Van Hove singularity away from the Fermi level,
thereby suppressing potential instabilities.
Finally, we investigate the altermagnetic phase in the broader class T$_{0.25}$MX$_2$,
finding that spin--orbit effects induce orbital antiferromagnetism with weak ferromagnetism or ferrimagnetism
depending on the Néel vector orientation.
Our results demonstrate that Ni-intercalated NbSe$_2$ provides a versatile platform to explore and tune
multiple competing magnetic phases that lie close in energy.

\end{abstract}


\maketitle
	
\section{Introduction}

The tunability of magnetic and electronic phases in transition metal dichalcogenides (TMDs) doped or intercalated with ionic liquids\cite{Zhang2022-rk}, through self-intercalation\cite{Wang2024-be}, or with molecules\cite{Scheidt2015-la} has attracted significant research interest due to its potential for spintronic applications and the generation of new exotic phases. 
Intercalation of transition metals into TMDs modifies the electronic structure, leading to changes in the exchange interactions that govern the magnetic ordering.
Focusing on the intercalation with magnetic transition metals, these materials exhibit diverse magnetic behaviors, including Weyl ferromagnetism\cite{PhysRevResearch.4.L042021}, chiral helimagnetism\cite{CAO2020100080}, altermagnetism\cite{devita2025opticalswitchinglayeredaltermagnet,candelora2025discoverymagneticfieldtunabledensitywaves} and spin glass states\cite{kong2023near,Maniv2021-vb}. This depends on the choice of dopant, concentration, and external factors such as strain and temperature\cite{He2024}.
The spatial distribution and local environment of the intercalate atoms play a crucial role in defining the resultant magnetic order. The degree of hybridization between transition metal d-states and chalcogen p-states can significantly impact the strength and the nature of magnetic interactions.\cite{Liu2024-du}
External stimuli, including electric gating\cite{yu2015gate}, pressure\cite{bai2022interlayer}, and strain engineering\cite{zhou2012tensile}, offer further tunability. Strain can enhance spin-orbit coupling (SOC) and modify superexchange interactions\cite{chen2015strain}, leading to novel magnetic phases. Similarly, electrostatic gating has been demonstrated to control magnetic order dynamically, enabling voltage-driven phase transitions\cite{khan2024phase}. These tunable properties make magnetic TMDs promising candidates for next-generation spintronic devices, including nonvolatile memories and logic applications.\cite{Nair2020-fo,PhysRevB.105.174404}

Several transition metal dichalcogenides host electronic instabilities such as charge density wave (CDW)\cite{Skolimowski2019-xx,PhysRevB.102.155115,Cossu2020-ei} and superconductivity\cite{PhysRevB.105.115110,PhysRevB.101.205149}.
In the ultrathin limit, NbSe$_2$ hosts superconductivity\cite{Ugeda2016,Xi2016-gc} and CDW\cite{rivano2025exploringchargedensitywaves}, which can interplay with different layered stacking orders\cite{PhysRevResearch.6.043111,PhysRevB.105.035119}, self-intercalation\cite{PhysRevB.108.L041405} and can host topological phases\cite{Chiu2025}.
One of the relevant ingredients for the electronic instabilities in NbSe$_2$ is the presence of Van Hove singularity (VHS) close to the Fermi level.\cite{PhysRevB.80.241108,BORISENKO2011562}
Several papers reported the successful synthesis of NbSe$_2$ intercalated with early 3d transition metals\cite{He2024}, while very few results have been reported for Ni intercalation, which is the focus of the present paper. When intercalating Ni in NbSe$_2$, the nearest neighbor (NN) magnetic exchange tends to be antiferromagnetic between the planes in the case of Ni${_{1/3}}$NbS$_2$\cite{PhysRevB.108.054418}. 
Ni${_{1/3}}$NbS$_2$ is altermagnetic\cite{Tenzin2025} with a critical temperature of 84 K\cite{PhysRevB.108.054418} and shares the same structure as the proposed altermagnet V$_{1/3}$NbS$_2$\cite{Ray2025,jungwirth2025symmetry,ghosh2025raman}.
The Ni$_{1/4}$NbSe$_2$ compound contains less Ni than Ni$_{1/3}$NbS$_2$ (there are 0.25 Ni per Nb atom versus 0.33 Ni per Nb atom), but we still expect the system to be magnetic. We expect a modest critical temperature, assuming a linear dependence on Ni composition and that the magnetism is relatively insensitive to whether S or Se is present.
The Ni atoms are positioned on top of the Nb atoms. The distribution of the Ni atoms is assumed to be uniform in density functional theory (DFT) calculations, while in experiment, there are vacancies that will add disorder to the magnetic structure, especially if we are far from the nominal stoichiometry\cite{erodici2023bridging}.
Recently, a compound with the same space group  \textcolor{red}{,}CoNb$_4$Se$_8$\textcolor{red}{,} was found to be altermagnetic due to the antiferromagnetic coupling between the Ni plane and antiferromagnetic coupling in the ab plane\cite{regmi2024altermagnetismlayeredintercalatedtransition,devita2025opticalswitchinglayeredaltermagnet}.
The only theoretical study about Ni$_{0.25}$NbSe$_2$ focuses on the small unit cell (two magnetic atoms) within standard DFT, finding that the system is ferromagnetic with a low magnetic moment\cite{RAMAKRISHNA2024172388}.We reproduce these results under the same conditions, but we expand the study to all possible magnetic phases.
Experimentally, a small amount of Ni intercalation in Ni$_{0.19}$NbSe$_2$ produces a low critical temperature and an electron pocket at the $\Gamma$ point\cite{Wadge2025submitted}, differently from the hole pocket of the pristine NbSe$_2$. The strong modifications of the Fermi surface, by the Ni atoms with respect to the pristine NbSe$_2$, disrupt the CDW\cite{wadge2025b}.  

\begin{figure}
    \centering
    \includegraphics[width=1\linewidth]{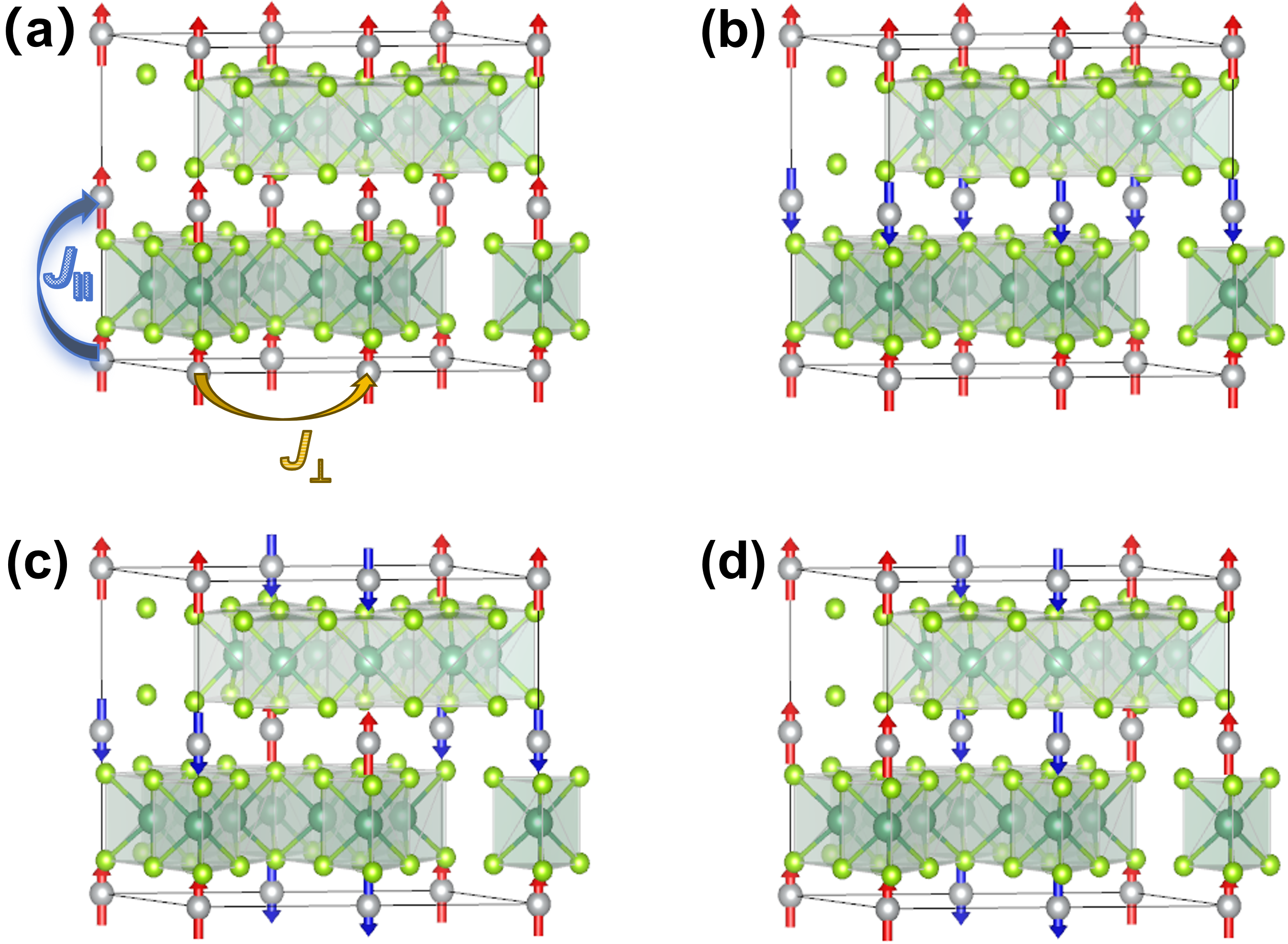}
    \caption{Four different possible magnetic structures of Ni$_{0.25}$NbSe$_2$. The magnetic configuration that breaks time-reversal symmetries are the ferromagnetic (FM) in panel (a) and the altermagnetic (AM) in panel (b). Once we double the supercell, we obtain two antiferromagnetic phases that we define as AFM1 in panel (c) and AFM2 in panel (d). The red (blue) arrows represent the atoms with the majority of spin-up (spin-down). The Ni, Nb and Se atoms are represented by grey, green and yellow balls, respectively. The NbSe$_6$ octahedra are in transparent green. The circular arrows in (a) represent the nearest-neighbor exchange interactions $J_{\parallel}$ and $J_{\perp}$.}
    \label{fig:1}
\end{figure}

In this paper, we study the magnetic properties of Ni$_{0.25}$NbSe$_2$ (also represented as NiNb$_4$Se$_8$ or Ni$_{\frac{1}{4}}$NbSe$_2$), including collinear and non-collinear phases, discussing the tunability of the magnetic properties in Ni intercalated
transition metal dichalcogenide NbSe$_2$. We will demonstrate that the Ni$_{0.25}$NbSe$_2$ transitions from a stripe antiferromagnetic phase to a ferromagnetic phase as a function of U, and discuss the possibility of a spin-glass phase. We will highlight the difference with the altermagnetic Ni$_{1/3}$NbSe$_2$ and estimate the magnetic exchanges. 
The paper is structured as follows: Section 2 discusses magnetic properties with SOC, while Section 3 investigates magnetic properties including relativistic effects. In the fourth and final Section, we draw our conclusions and outlook.

\begin{table}
\begin{tabular}{ccccc}
  \Xhline{1pt}
  Atom & WP & \( x \) & \( y \) & \( z \) \\
  \hline
  Ni & \( 2a \) & 0 & 0 & 0 \\
  Nb(1) & \( 2b \) & 0 & 0 & $\frac{3}{4}$ \\
  Nb(2) & \( 6h \) & 0.01949 & 0.50975 & $\frac{3}{4}$ \\
  Se(1) & \( 12k \) & 0.83145 & 0.66290 & 0.88328 \\
  Se(2) & \( 4f \) & $\frac{2}{3}$ & $\frac{1}{3}$ & 0.39321 \\
  \Xhline{1pt}
\end{tabular}
\caption{Atomic positions of the primitive cell of the NM magnetic structure calculated by DFT+U (U=0 eV), WP is the abbreviation of Wyckoff positions. The space group is no. 194.}
\label{NMpositions}
\end{table}

\section{Electronic and Magnetic properties without SOC}

In this section, we report the structural properties and the study of the collinear magnetic phases as a function of the electronic correlations for Ni$_{0.25}$NbSe$_2$.

\subsection{Structural relaxation and magnetic ground state}

As it was confirmed from experimental studies for CoNb$_4$Se$_8$.\cite{regmi2024altermagnetismlayeredintercalatedtransition}, the position of the magnetic atoms tends to be on top of Nb in transition metals intercalated NbSe$_2$ with Ni$_{0.25}$NbSe$_2$ stoichiometry. 
The experimental lattice constants of pristine bulk NbSe$_2$ were 3.44445 {\AA} and 12.54527 {\AA}. The experimental lattice constants for the equivalent cell in Ni$_{0.25}$NbSe$_2$ are 3.455 {\AA} and 12.3238 {\AA}, with a reduction of the in-plane lattice constant and an increase of the out-of-plane lattice constant.\cite{wadge2025b}
The unit cell of Ni$_{0.25}$NbSe$_2$ can be obtained from NbSe$_2$ by doubling along the a and b-axis using a 2$\times$2$\times$1 supercell; this supercell contains two intercalated Ni atoms and allows us to simulate the magnetic phases in Fig. \ref{fig:1}(a) and \ref{fig:1}(b). By using a 4$\times$2$\times$1 supercell containing 4 magnetic atoms, we can simulate the magnetic phases reported in Fig. \ref{fig:1}(c) and \ref{fig:1}(d), which have two Ni atoms per plane. The Ni atoms form a triangular lattice in the ab plane, while they are on top of each other regarding the out-of-plane connectivity. In all these 3d-intercalated transition metal dichalcogenides, we have both transition metals surrounded by 6 chalcogen atoms forming octahedra. 

We use the unit cell to simulate the FM and AM phases, while we use a supercell for the AFM1 and AFM2 phases. 
We report the Brillouin zone (BZ) for the two space groups considered for this compound: P6$_3$/$mmc$ (no. 194) for the unit cell and P2$_1$ (no. 4) for the supercell. We perform structural relaxation, constraining the volume to the experimental values. The outputs for the internal degrees of freedom are reported for the nonmagnetic phase (NM), the FM and the AM phase in Tables \ref{NMpositions}, \ref{FMpositions} and \ref{AMpositions}, respectively. 
We note that the internal degrees of freedom barely change from one magnetic phase to another. It is worth mentioning that we have equivalent Nb and Se atoms. The Nb atoms at the 2b Wyckoff positions and the Se atoms at the 4f positions are the closest to the Ni atoms, while the Nb atoms at the 6h positions and the Se atoms at the 12k positions are the farthest.

\begin{table}
\begin{tabular}{ccccc}
  \Xhline{1pt}
  Atom & WP & \( x \) & \( y \) & \( z \) \\
  \hline
  Ni & \( 2a \) & 0 & 0 & 0 \\
  Nb(1) & \( 2b \) & 0 & 0 & $\frac{3}{4}$ \\
  Nb(2) & \( 6h \) & 0.01935 & 0.50968 & $\frac{3}{4}$ \\
  Se(1) & \( 12k \) & 0.83135 & 0.66271 & 0.88330 \\
  Se(2) & \( 4f \) & $\frac{2}{3}$ & $\frac{1}{3}$ & 0.39298 \\
  \Xhline{1pt}
\end{tabular}
\caption{Atomic positions of the primitive cell of FM magnetic structure calculated by DFT+U (U=0 eV), WP is the abbreviation of Wyckoff positions. The space group is no. 194.}
\label{FMpositions}
\end{table}

\begin{table}

\begin{tabular}{ccccc}
  \Xhline{1pt}
  Atom & WP & \( x \) & \( y \) & \( z \) \\
  \hline
  Ni & \( 2a \) & 0 & 0 & 0 \\
  Nb(1) & \( 2b \) & 0 & 0 & $\frac{3}{4}$ \\
  Nb(2) & \( 6h \) & 0.01968 & 0.50984 & $\frac{3}{4}$ \\
  Se(1) & \( 12k \) & 0.83092 & 0.66185 & 0.88331 \\
  Se(2) & \( 4f \) & $\frac{2}{3}$ & $\frac{1}{3}$ & 0.39262 \\
  \Xhline{1pt}
\end{tabular}
\caption{Atomic positions of the primitive cell of AM magnetic structure calculated by DFT+U (U=3 eV), WP is the abbreviation of Wyckoff positions. The space group is no. 194.}
\label{AMpositions}
\end{table}

\begin{figure}
    \centering
    \includegraphics[width=1\linewidth]{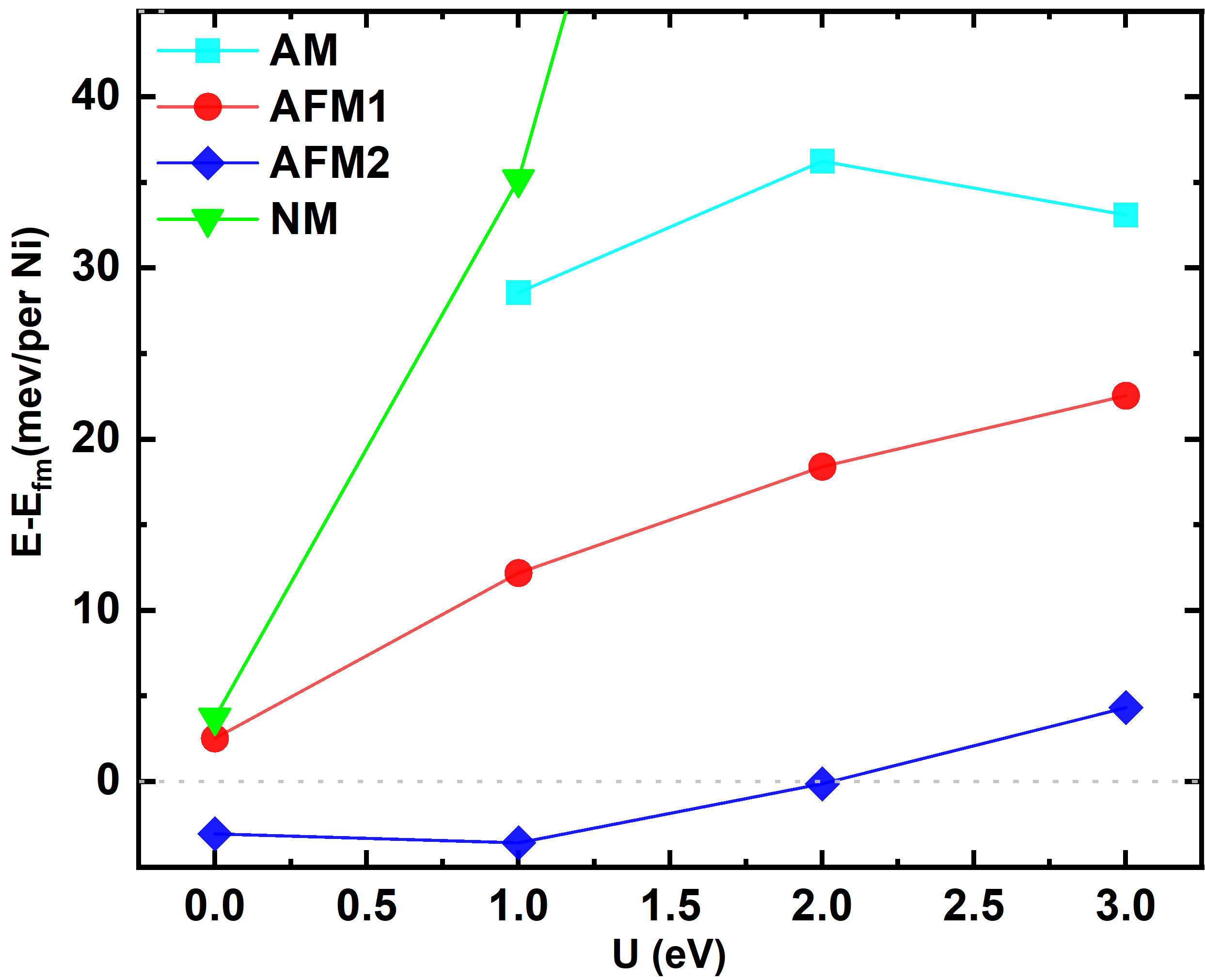}
    \caption{Energy difference of AM, AFM1, AFM2 configurations relative to FM configuration per Ni as a function of U ranging from 0 to 3 eV. These data refer to the calculations performed without SOC. The energy difference of the NM configuration is out of the scale, since it reaches the values of 98 and 188 meV for U=2 eV and 3 eV, respectively.}
    \label{fig:2}
\end{figure}

Given that the supercell contains 4 atoms, we can consider the ferromagnetic phase (FM) and 3 independent magnetic phases with zero magnetization. The real space pictures of these magnetic phases are reported in Fig. \ref{fig:1}.
The FM phase is reported in Fig. \ref{fig:1}(a). We will demonstrate that the magnetic phase with in-plane ferromagnetic coupling and out-of-plane antiferromagnetic coupling is altermagnetic, we name it AM, and we report it in  Fig. \ref{fig:1}(b). The other two magnetic configurations are antiferromagnetic with Kramers' degeneracy; they are illustrated in Fig. \ref{fig:1}(c-d) and we define them as AFM1 and AFM2, respectively.
AFM1 hosts both in-plane and out-of-plane antiferromagnetic coupling, while AFM2 hosts ferromagnetic coupling out-of-plane and antiferromagnetic coupling in-plane.
Within the single layer, AFM1 and AFM2 have the same magnetic order, which is usually named as stripe phase magnetic order. 
We present in Fig. \ref{fig:2} the energy differences between these 4 magnetic configurations. 
In our theoretical results, even for U=0, the ground state is magnetic. Therefore, we expect the system to be experimentally magnetic at low temperatures in the case of nominal stoichiometry and ideal crystal structure.
Note that the AM phase results in Fig. \ref{fig:2} are reported starting from U=1 eV, since at U=0 eV, the AM phase tends to converge to a nonmagnetic state with vanishing magnetic moments after structural relaxation of the internal degrees of freedom.
As shown in Fig. \ref{fig:2}, the FM and AFM2 configurations are the most stable phases. The ground state depends on the value of the electronic correlations; there is a critical value U$_C$ at which we have a change of the ground state.
The AFM2 phase is the ground state for U$<$U$_C$ while the ferromagnetic phase is the ground state for U$>$U$_C$. In the non-relativistic case, we estimate U$_C\approx$2.1 eV. As we will see in the next sections, the system is metallic with a large density of states at the Fermi level. The metallicity strongly screens the Coulomb repulsion, lowering its value. 
Since the experimental magnetic moments of metallic Ni$_{1/3}$NbS$_2$ are reproduced using GGA+U with U$\approx$1 eV\cite{PhysRevB.108.054418,Tenzin2025}, we infer that AFM2 is likely the ground state of Ni$_{0.25}$NbSe$_2$ under ideal stoichiometry and ordered Ni distribution.

\begin{figure}
    \centering
    \includegraphics[width=1\linewidth]{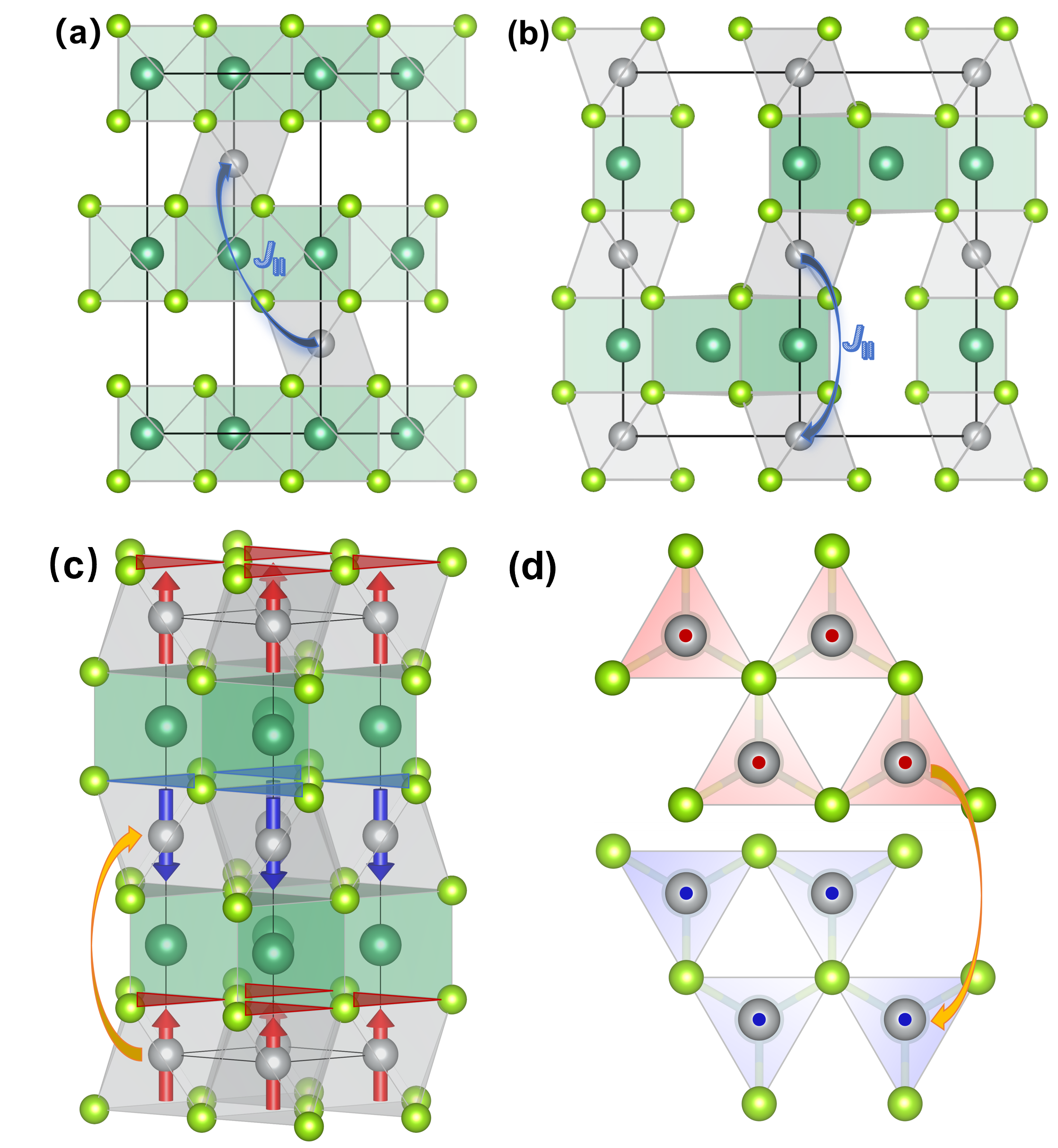}
    \caption{The out-of-plane magnetic coupling J$_{||}$ has different connectivity in (a) Ni$_{1/3}$NbS$_2$ with respect to (b) Ni$_{0.25}$NbSe$_2$. Due to this different connectivity, the coupling of Ni$_{1/3}$NbS$_2$ is antiferromagnetic, while the coupling of Ni$_{0.25}$NbSe$_2$ is ferromagnetic. (c) Side and (d) top view of NiNbSe$_2$. Red and blue arrows represent the spin-up and spin-down sites. We highlight the C$_2$ rototranslational symmetry that connects the spin-up and spin-down states, generating alternagnetism due to the absence of a translational symmetry. The C$_2$ spin rotation is represented by the orange arrow. Ni, Nb, and chalcogen atoms are represented by gray, green and yellow atoms, respectively.}
    \label{fig:comparison}
\end{figure}

The out-of-plane ferromagnetic coupling between the Ni atoms in Ni$_{0.25}$NbSe$_2$ differs from the out-of-plane antiferromagnetic coupling found in Ni$_{1/3}$NbS$_2$ despite the two-dimensional material hosts being isoelectronic. The difference lies in the connectivity between the Ni atoms. The Ni atoms of different layers in Ni$_{0.25}$NbSe$_2$ are one on top of each other as reported in Fig. \ref{fig:comparison}(b), while they are not in Ni$_{1/3}$NbS$_2$ as we can see in Fig. \ref{fig:comparison}(a).
In Fig. \ref{fig:comparison}(c) and Fig. \ref{fig:comparison}(d), we report the 
schematic view of the crystal and magnetic structure of the altermagnetic phase of Ni$_{0.25}$NbSe$_2$ in the y–z and x–y planes, respectively. The green octahedra are the NbSe$_6$ octahedra. The red and blue shadings mark the NiSe$_6$ octahedra around the Ni sites with opposite spins, which are related by the twofold spin rotation combined with the sixfold crystal rotation and half-unit cell translation along the z-axis. These are the same symmetries involved in the NiAs crystal structure and produce non-relativistic spin-splitting in the band structure\cite{Smejkal22,Cuono23orbital,Cuono23EuCd2As2,PhysRevB.108.115138}; therefore, the non-relativistic band structure will have the same symmetries as MnTe and CrSb with the spin-momentum locking of the kind B-4\cite{PhysRevX.12.031042}. However, we will show later that the weak ferromagnetism will behave differently due to the presence of additional atoms, which lowers the symmetry of the magnetic space group.

\subsection{Band structure}

\begin{figure}
    \centering
    \includegraphics[width=1\linewidth]{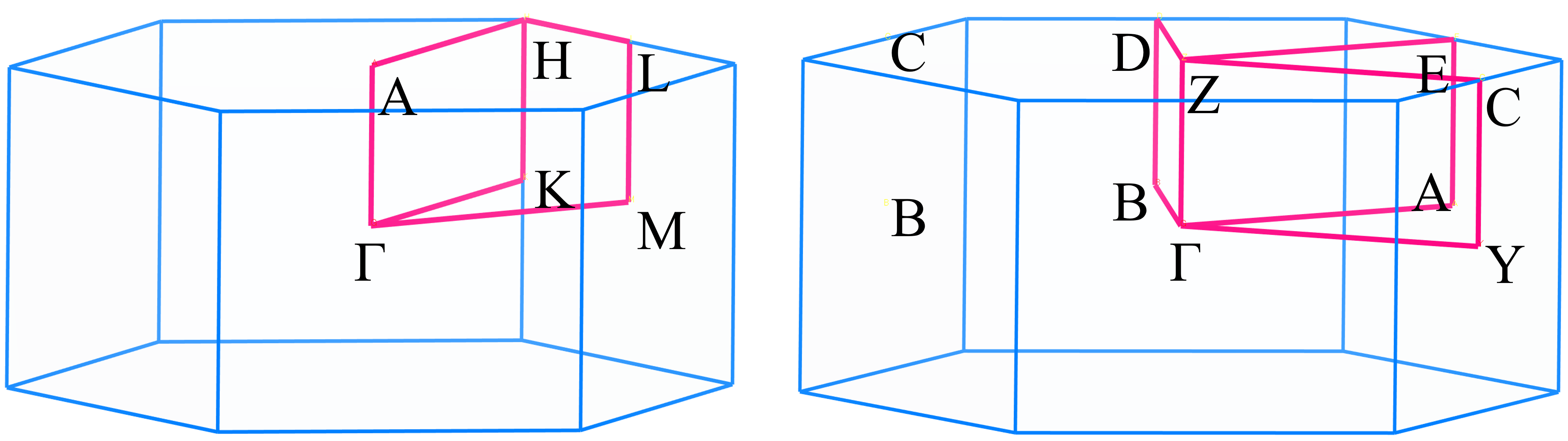}
    \caption{Brillouin zone for the space group no.194 and no.4 of Ni$_{0.25}$NbSe$_2$. Left panel: Brillouin zone of the unit cell with space group no. 194. Right panel: Brillouin zone of the supercell with space group no. 4. The high-symmetry points E and D are not equivalent because of the doubling of the unit cell along one direction. }
    \label{fig:Brillouin}
\end{figure}

\begin{figure}
    \centering
    \includegraphics[width=1\linewidth]{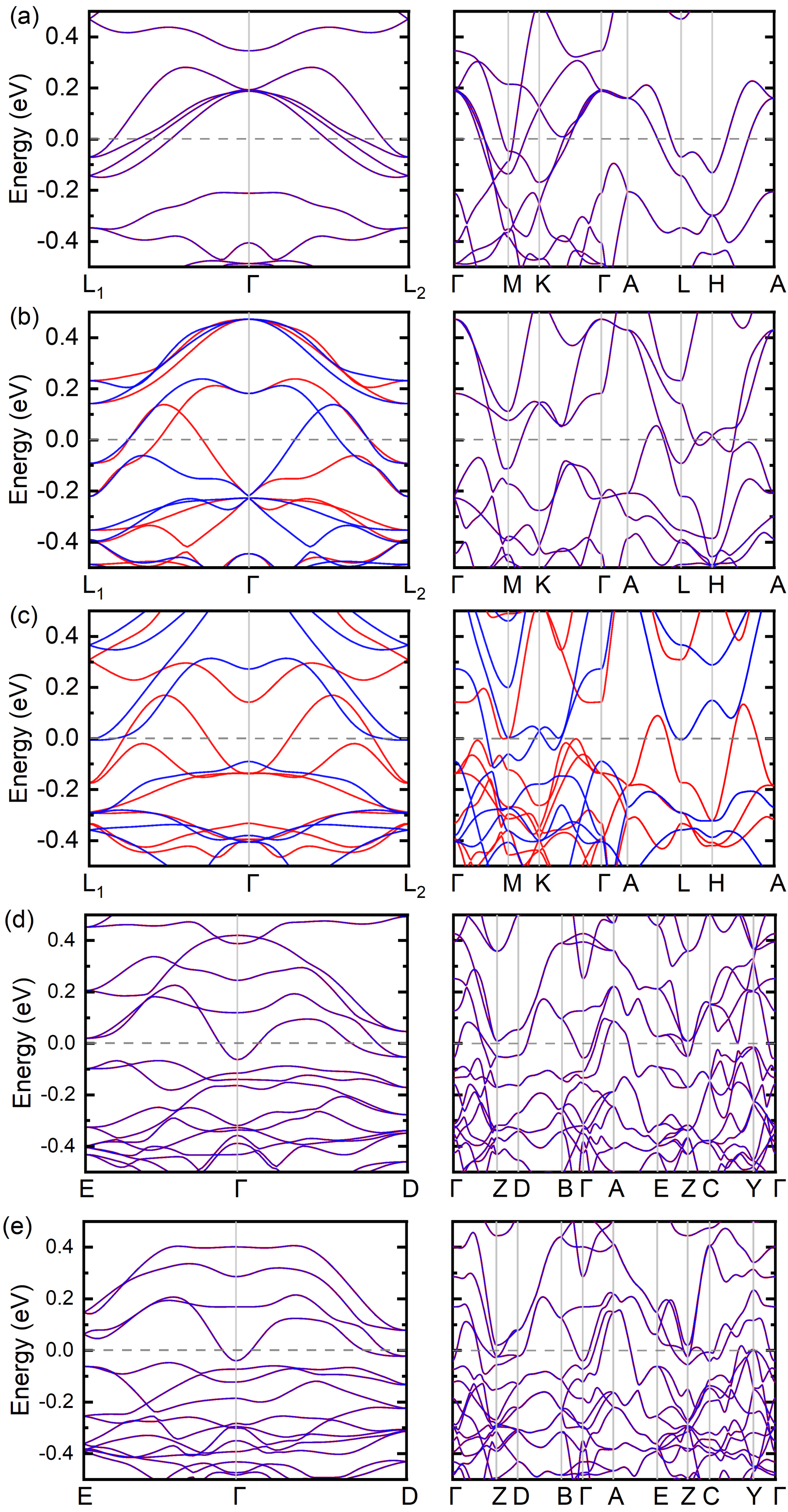}
    \caption{Electronic band structure of Ni$_{0.25}$NbSe$_2$ for different magnetic states: (a) NM (U=0), (b) AM (U=2 eV), (c) FM (U=3 eV), (d) AFM1 (U=2 eV), and (e) AFM2 (U=2 eV). Red/blue denote spin-up/down (merging into purple under Kramers' degeneracy). Left: zoom near $\Gamma$; right: full $k$-path. AFM2 and FM are shown at their ground-state $U$ values.}
    \label{fig:band_structure}
\end{figure}

We calculate the band structure for the 4 magnetic phases and the nonmagnetic phase. The k-paths for the two space groups are reported in Fig. \ref{fig:Brillouin}, while the band structures are reported in Fig. \ref{fig:band_structure}. 
We plot the results of the nonmagnetic phase at U=0 in Fig. \ref{fig:band_structure}(a), while we plot in Fig. \ref{fig:band_structure}(c) the band structure of the ferromagnetic phase for U=3 eV when it is the ground state. For the other phases, we plot the band structures at U=2 eV in Fig. \ref{fig:band_structure}(b,d,e). We report the band structure for the value of U for which the relative phases are the ground state. There is a negligible difference in the band structure between U=2 and U=3 eV for the metallic phases where the Coulomb repulsion is screened; therefore, for small variations of U, the band structures will look similar.

We observe the non-relativistic spin-splitting in the FM and AM phases, while AFM1 and AFM2 host Kramers degeneracy. The AM band structure in Fig. \ref{fig:band_structure}(b) exhibits a non-relativistic spin-splitting up to 0.2 eV along L$_1$-$\Gamma$-L$_2$ with a g-wave spin-momentum locking.
While the relaxed atomic positions are basically not affected by the magnetism, the band structure strongly changes across the magnetic phases.  
For instance, we can focus on the $\Gamma$ point.
In bulk pristine NbSe$_2$, at the $\Gamma$ point, there are bands with negative effective masses (hole pockets) in the low-energy band structure\cite{rossnagel2001fermi}\cite{rahn2012gaps}\cite{johannes2006fermi}. Unlike pristine NbSe$_2$, all the magnetic phases feature an electron pocket with positive effective mass at the $\Gamma$ point in its low-energy band structure. The Ni intercalation changed the character of the carriers at the $\Gamma$ point from hole to electrons. 
To gain further insight, we focus on AFM2, which is the ground state at low Coulomb repulsion. We examined the orbital character of the electron band in the low-energy range at the $\Gamma$ point for the AFM2 phase. We find that the electron band below the Fermi level is mainly composed of the 3z$^2$-r$^2$ electrons of Nb with additional Nb components of xy and x$^2$-y$^2$ plus contribution from the same orbitals on the Ni atoms.
In the FM phase, the electron pocket is composed of the same orbitals but only from Nb atoms.

\begin{figure*}
    \centering
    \includegraphics[width=1\linewidth]{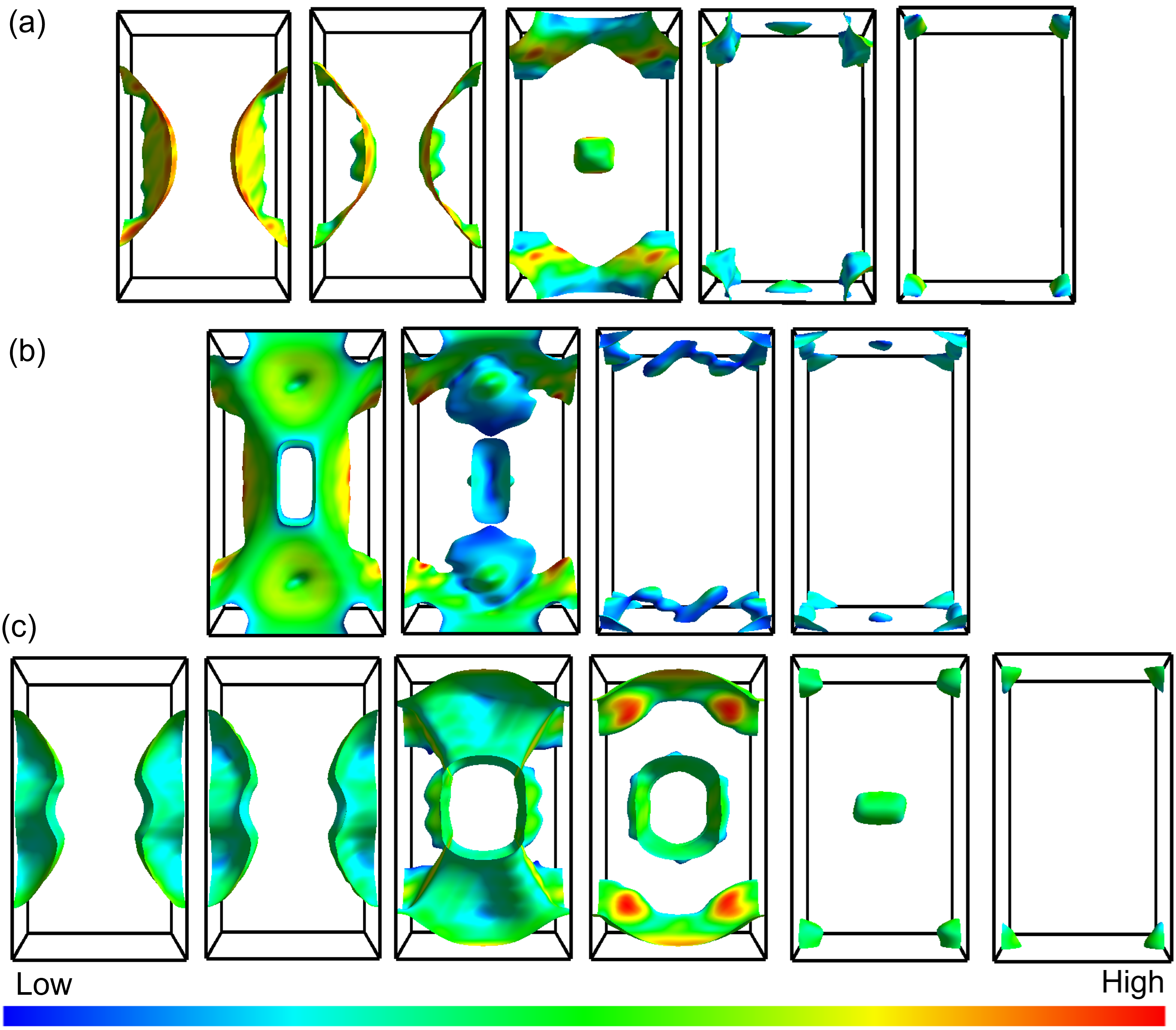}
    \caption{Fermi surface calculated within GGA+U at U=2 eV for the : (a) AFM2, (b) spin-down of FM and (c) spin-up of FM. The BZ has a parallelepiped shape, due to the expansion of the unit cell. The $\Gamma$ point lies in the middle of the BZ, and several figures for each case indicate different bands. The Fermi surface was shown in a top view with a perspective image. The color legend represents low and high Fermi velocities on the Fermi surfaces}.
    \label{Fermi_surfaces}
\end{figure*} 
\begin{figure}
    \centering
    \includegraphics[width=1\linewidth]{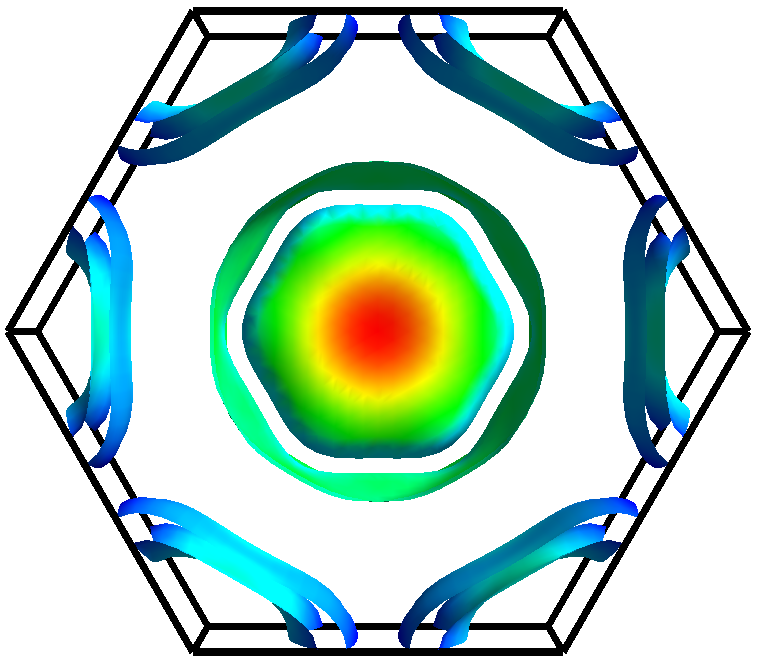}
    \caption{Three sheets of the Fermi surfaces for the pristine NbSe$_2$ including spin-orbit coupling. These Fermi surfaces show doubly nested hole pockets around the $\Gamma$ point\cite{johannes2006fermi}. The colors represent low and high Fermi velocities on the Fermi surfaces as in the previous figure.}
    \label{fig:7}
\end{figure}

\subsection{Fermi surface}

We report the Fermi surfaces in the first BZ for the AFM2 and FM phases in Fig. \ref{Fermi_surfaces}. To compare the two phases, we use for both cases a 4$\times$2$\times$1 supercell (see Appendix). In this particular supercell, we do not find the hexagonal BZ of the pristine NbSe$_2$, but the BZ is a parallelepiped. 
The AFM2 phase hosts 5 Fermi surfaces as shown in Fig. \ref{Fermi_surfaces}(a). In the FM phase, spin splitting lifts the degeneracy resulting in distinct spin-up and spin-down sheets that are plotted in Fig. \ref{Fermi_surfaces}(b) and \ref{Fermi_surfaces}(c), respectively. There are 4 Fermi surfaces for the spin-up and 6 for the spin-down channel.
In all phases, we have hole- and electron-pockets. We can observe significant differences between the Fermi surfaces of the AFM2 and the FM phases, where it is more fragmented in the AFM2 phase, where smaller pockets appear at the boundaries. On the other hand, the Fermi surface of the FM phase shows smoother and more connected sheets.


Unlike the Fermi surface of NbSe$_2$ presented in Fig. \ref{fig:7} which exhibits doubly nested hole pockets around the $\Gamma$ point\cite{rossnagel2001fermi}\cite{johannes2006fermi}, significant electron pockets emerge in both the AFM2 and FM phases upon Ni intercalation in agreement with band structure calculations. Along with hole-pockets present in all figures, the electron pockets at $\Gamma$ are visible mainly in the third panel of Fig. \ref{Fermi_surfaces}(a), in the second panel of Fig. \ref{Fermi_surfaces}(b), and in the fifth panel of Fig. \ref{Fermi_surfaces}(c). The characteristic isolated triangular pockets observed in pristine NbSe$_2$ vanish, and in certain bands, the Fermi surface evolves into a fully complex interconnected topology. Furthermore, compared to NbSe$_2$, the Fermi surface dispersion along the $k_z$ direction is markedly enhanced. The latter is due to the Ni atoms, which work as bridges between the NbSe$_2$ layers, making the system more three-dimensional. Both the FM and AFM2 phase have an electron pocket at the $\Gamma$ point with the FM phase showing a slightly larger pocket. Regarding the size of the Fermi sheets, the FM phase exhibits broader sheets than the AFM phase, which can be attributed to the enhanced metallicity that is usually observed in the ferromagnetic phases compared to antiferromagnetic ones.


\subsection{Nearest-neighbor spin exchange coupling $J$}

The nearest-neighbor spin exchange interaction parameter $J$ can be determined by mapping the system onto the Heisenberg model with effective spin S=1. The corresponding spin Hamiltonian can be expressed as\cite{White2006}:
\begin{eqnarray}
    H = \sum_{{i},{j>i}}J_{ij}\vec{S}_{i}{\cdot}\vec{S}_{j}
\end{eqnarray}
Here, $S$ represents the total spin vector of the Ni atoms. For Ni$_{0.25}$NbSe$_2$, Ni exists in an electronic configuration close to the Ni$^{2+}$(3d$^8$) state. This means that the triply degenerate t${_{2g}}$ states are fully occupied, while the doubly degenerate e{$_g$} states accommodate two electrons. Consequently, Ni adopts a low-spin state with a total spin of $S$=1. The magnetic moments extracted from the ground state phase AFM2 range from 0.436 $\mu_B$ at U=0 to 0.975 $\mu_B$ at U=3 eV. 
Increasing the value of U will give a magnetic moment which is above 1 $\mu_B$ for the Ni atoms; therefore, the system will adopt a high-spin configuration with the t$_{2g}$ orbitals of the minority channel remaining partially occupied.
In the ferromagnetic phase, the Nb atoms exhibit proximitized magnetization, with a magnetic moment of 0.09 $\mu_B$ at the 2b Wyckoff position and 0.03 $\mu_B$ at the 6h Wyckoff position for U=2 eV. In all other cases, the magnetization on the Nb atoms is zero due to the compensation arising from the hybridization with the magnetic atoms of opposite spins. 

We aim to calculate the nearest-neighbour magnetic exchange interactions: one in the out-of-plane direction and one in-plane. Considering the nearest-neighbor exchange interactions, we define the out-of-plane exchange constant as $J_{\parallel}$ and the in-plane exchange constant as $J_{\perp}$. To simplify the calculations, we map the Heisenberg model on spins with $S$=1. In this framework, the total energies of these magnetic structures can be expressed as follows:

\begin{eqnarray}\label{Jformula1}
    E_{FM}=E_0+4J_{\parallel}+12J_{\perp}\\ \label{Jformula2}
    E_{AFM1}=E_0-4J_{\parallel}-4J_{\perp}\\ \label{Jformula3}
    E_{AFM2}=E_0+4J_{\parallel}-4J_{\perp}\\    
    E_{AM}=E_0-4J_{\parallel}+12J_{\perp}
\end{eqnarray}

While $E_0$ is the nonmagnetic total energy of the system. The same equations are valid with and without SOC. To determine the nearest-neighbor exchange parameter \( J \) from DFT calculations, we employ the Heisenberg Hamiltonian and establish a relationship between the total energies of only three distinct magnetic configurations: 
From the previous set of equations, we will consider the first three equations (\ref{Jformula1})-(\ref{Jformula3}) to derive J$_{\parallel}$ and J$_{\perp}$.
These equations will be used in the next Section to derive the magnetic exchanges J$_{\parallel}$ and J$_{\perp}$ in the non-relativistic and relativistic cases. 

\subsection{Magnetic coupling at different Ni concentrations}

\begin{figure}
    \centering
    \includegraphics[width=1\linewidth]{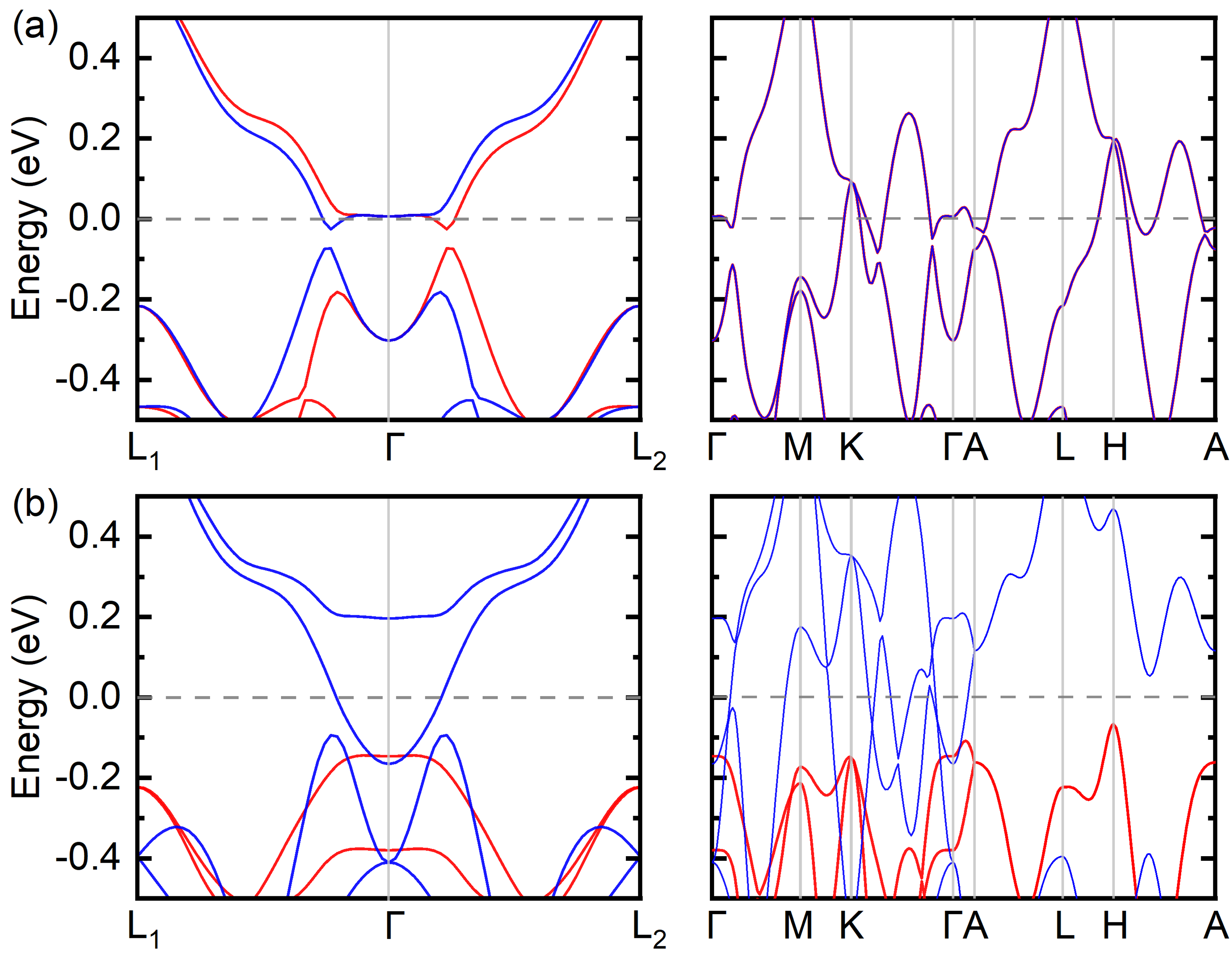}
    \caption{Non-relativistic band structure of NiNbSe$_2$ for the (a) altermagnetic and (b) ferromagnetic phase at U=2 eV. On the left panel is the magnification close to the $\Gamma$ point, while the right panel is the band structure across the entire k-path. We observe non-relativistic spin-splitting along the L$_1$-$\Gamma$-L$_2$ path.}
    \label{bandstructure_ninbse2}
\end{figure}

\begin{figure}
    \centering
    \includegraphics[width=1\linewidth]{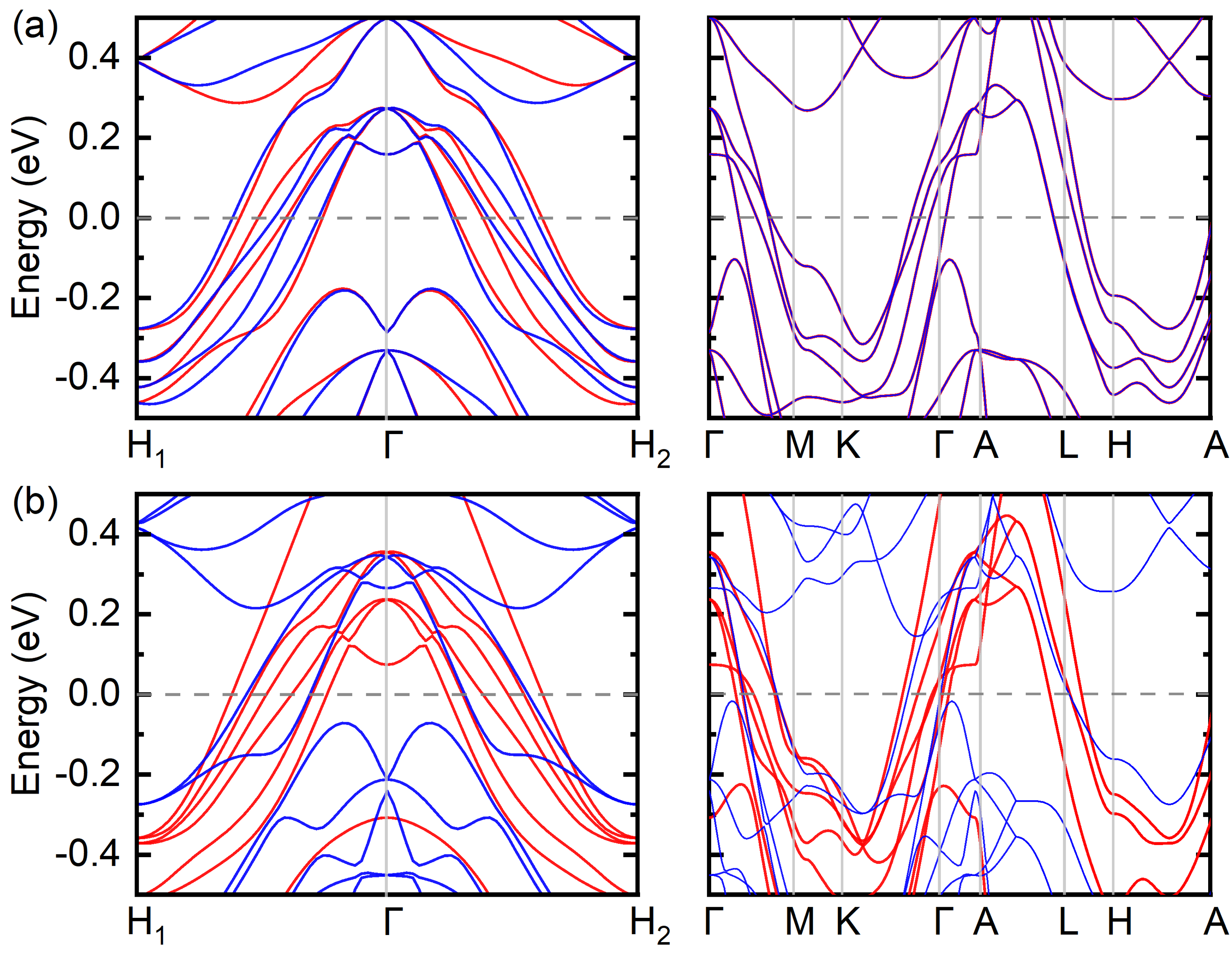}
    \caption{Non-relativistic band structure of Ni$_{_{1/3}}$NbSe$_2$ for the (a) altermagnetic and (b) ferromagnetic phase at U=2 eV. On the left panel is the magnification close to the $\Gamma$ point, while the right panel is the band structure across the entire k-path. We observe non-relativistic spin-splitting along the H$_1$-$\Gamma$-H$_2$ path.}
    \label{bandstructure_ni033nbse2}
\end{figure}

We calculate the magnetic coupling J$_{||}$  for two other Ni concentrations of Ni$_x$NbSe$_2$: Ni$_{_{1/3}}$NbSe$_2$ and NiNbSe$_2$, both of which have 2 Ni atoms per unit cell. The lattice constants which were used for Ni$_{0.25}$NbSe$_2$ were kept the same for different concentrations, so that the Nb-Nb distances are constant for both in-plane and out-of-plane. \textcolor{red}{However}, the Ni valence state and the connectivity between the Ni atoms change. In both cases, the system exhibits a ferromagnetic configuration, whereas the configuration with antiferromagnetic coupling breaks time-reversal symmetry, resulting in an altermagnetic phase. We report the band structure of the altermagnetic and ferromagnetic phases of NiNbSe$_2$ in Figs. \ref{bandstructure_ninbse2}(a) and 
\ref{bandstructure_ninbse2}(b), respectively.
We report the band structure of the altermagnetic and the ferromagnetic phases of Ni$_{_{1/3}}$NbSe$_2$ in Figs. \ref{bandstructure_ni033nbse2}(a) and 
\ref{bandstructure_ni033nbse2}(b), respectively.

Regarding the magnetic ground state, we start from the compound NiNbSe$_2$ and we get the ferromagnetic J$_{||}$, confirming that J$_{||}$ is ferromagnetic when the Ni atoms are on top of each other. In the ferromagnetic phase, the system is half metallic and the magnetic moment is 1 $\mu_B$. The Ni atoms are in a d$^9$ magnetic configuration, corresponding to an oxidation state of +1. As a consequence, the Nb is in oxidation state +3, making the Nb a 4d$^2$ system similiar to the insulating MoSe$_2$. For this reason, the Nb is depleted from the Fermi level, where only the minority spins of the Ni atoms contribute.
Ni would tend to have an oxidation state +2 due to the octahedral crystal field which raises the energy of the 4s level. However, if Ni is +2, Nb would not be able to be +4 as in pristine NbSe$_2$. Therefore, in the stoichiometry Ni$_x$NbSe$_2$, the Ni atoms have valence +2-$\delta$, where $\delta$ would tend to increase as the Ni concentration x increases.\cite{PhysRevB.109.085135}  
Since for x=1, we have $\delta$=1, and for x=0 we have $\delta$=0, and assuming that $\delta$ will depend linearly on the Ni concentration, we can establish that $\delta\approx{x}$. Therefore, in Ni$_x$NbSe$_2$ the oxidation state is Ni$^{+2-x}$, also, the results of the charge at x=1/4 and 1/3 confirms this trend in the first approximation.

\begin{figure}
    \centering
    \includegraphics[width=1\linewidth]{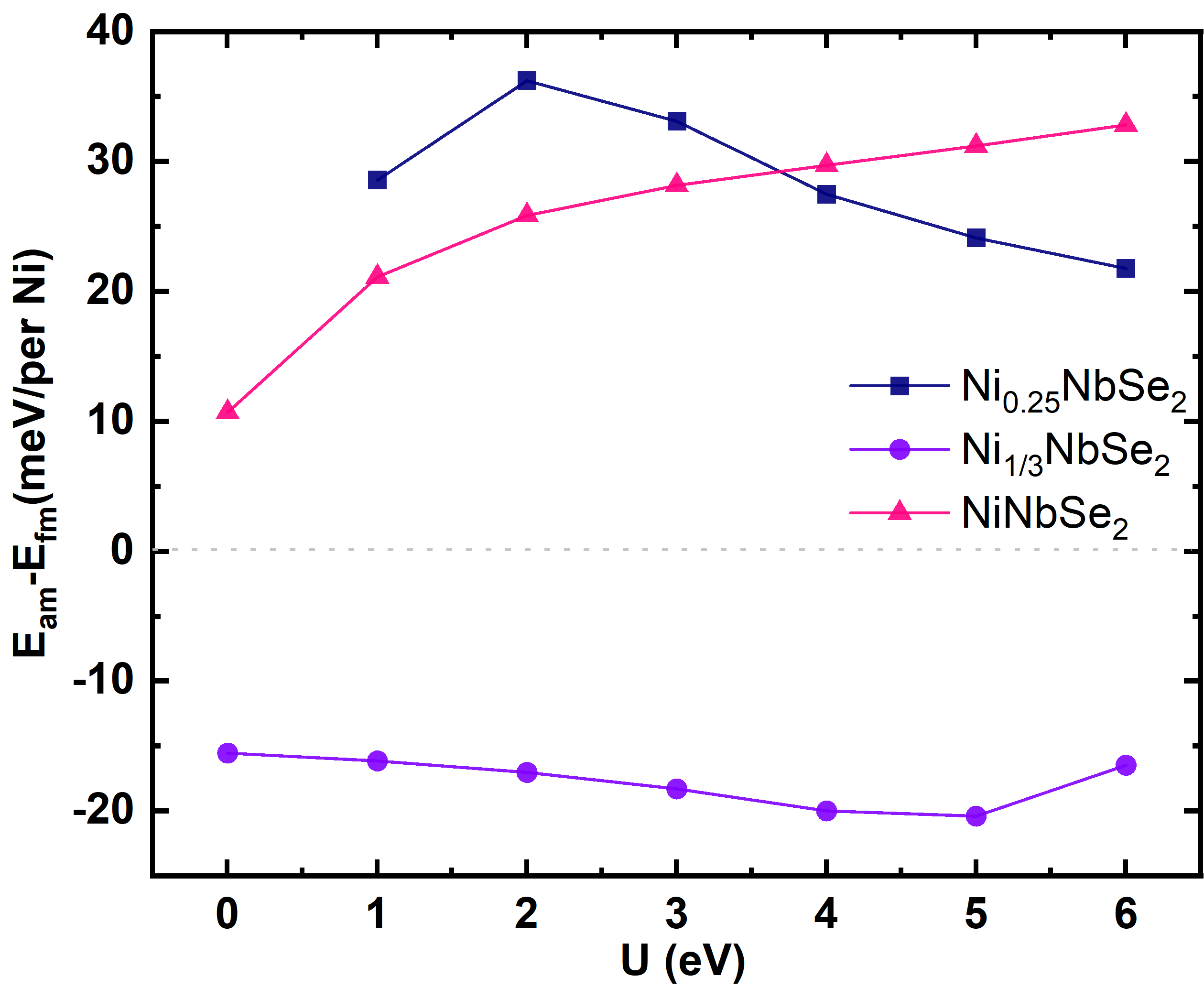}
    \caption{Energy difference between the AM and FM configurations of Ni$_x$NbSe$_2$ (x=0.25, 1/3, 1) as a function of U ranging from 0 to 6 eV within GGA + U. When the Ni atoms are one on top of each other, the coupling J$_{||}$ is ferromagnetic; otherwise it is antiferromagnetic. }
    \label{nix_energydiff}
\end{figure}

We compare the energies of the FM vs AM for the three cases of Ni$_{0.25}$NbSe$_2$, Ni$_{_{1/3}}$NbSe$_2$ and  NiNbSe$_2$. The results are reported in Fig. \ref{nix_energydiff}
As previously discussed, the position of Ni atoms in the intercalation layer significantly influences the connectivity. In the case of Ni$_{0.25}$NbSe$_2$ and NiNbSe$_2$, their identical connectivity leads both systems to favor a FM ground state. In contrast, Ni$_{1/3}$NbSe$_2$ exhibits a more stable AM configuration, with an energy at least 30 meV/Ni lower than that of the AM states of Ni$_{0.25}$NbSe$_2$ and NiNbSe$_2$ at U=2 eV. Although not yet experimentally confirmed, our calculations suggest that Ni$_{1/3}$NbSe$_2$ is a strong candidate material for exhibiting the novel altermagnetic phase.

For the compound NiNbSe$_2$, we have also performed metaGGA calculations to test the stability of the J$_{||}$ in this class of materials. The employed metaGGA functionals are 
SCAN\cite{sun2015}, RSCAN\cite{bartok2019} and R2SCAN\cite{furness2020}. We obtain an energy difference E$_{AM}$-E$_{FM}$ of 32 meV per Ni atom for all the functionals. This result confirms that when the Ni atoms are one on top of each other, the magnetic coupling between them is strongly ferromagnetic.
Also in metaGGA, we obtain a half-metallic system for the ferromagnetic phase of NiNbSe$_2$.

\section{Electronic and magnetic properties with SOC: role of the MCA}

In this section, we report the study of the magnetic phases as a function of the electronic correlations for Ni$_{0.25}$NbSe$_2$ in the relativistic case. Compared to the previous section, we include the investigation of the magnetocrystalline anisotropy and the study of the non-collinear 120$^{\circ}$ phase. Given the triangular lattice in the plane, if the system is antiferromagnetic in the plane, the magnetic system could become frustrated and the magnetism will become noncollinear\cite{PhysRevMaterials.4.074002}.

\subsection{Magnetocrystalline anisotropy of the ferromagnetic phase}

We calculate the magnetocrystalline anisotropy energy for the ferromagnetic phase at U=0 and U=3 eV. The easy axis lies along the z-axis. We can write the magnetocrystalline anisotropy as 
\begin{equation}
    E=K_2\sin^2{\theta}
\end{equation}
where $\theta$ is the polar angle of the spins\cite{Blundell2001}. We obtain that the magnetocrystalline anisotropy in the ferromagnetic phase is K$_2$=0.6 meV per atom for U=0 and K$_2$=0.9 meV per atom for U=3 eV. This relatively large value arises from the system’s strong crystalline anisotropy due to its layered crystal structure\cite{baranov2011magnetism}.

\begin{figure}
    \centering
   \includegraphics[width=1\linewidth]{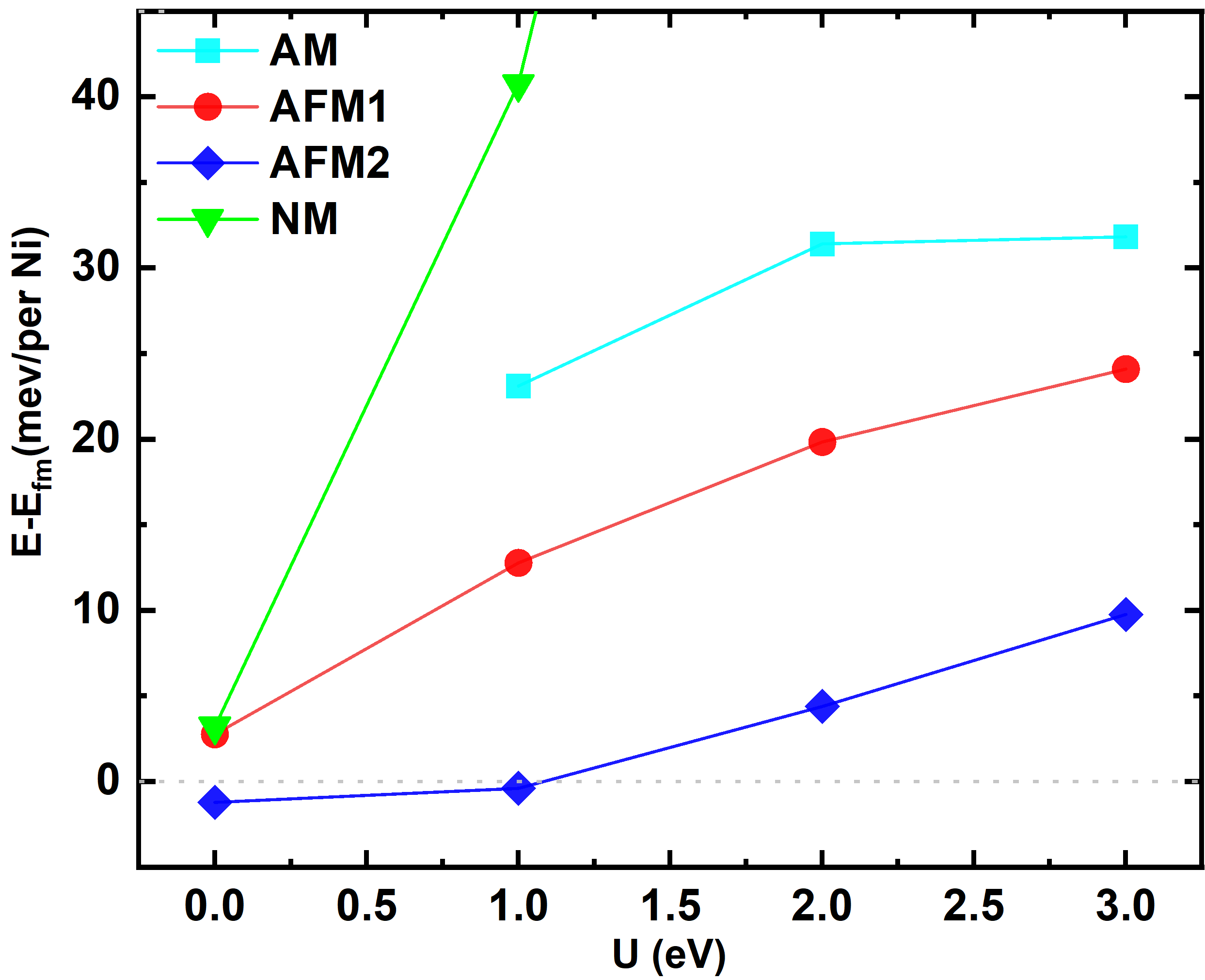}
    \caption{Energy difference of AM, AFM1, AFM2 configurations with respect to the FM configuration per Ni as a function of U ranging from 0 eV to 3 eV. As in the calculations without SOC, the results for the NM configuration when U exceeds 1 eV is not shown in the figure because it is several orders of magnitude larger than the other values.}
    \label{fig:energysoc}
\end{figure} 

\subsection{Magnetic ground state with SOC: NN magnetic couplings}

Using the same strategy used in the previous section, we have calculated the energy differences of the four collinear magnetic phases in the presence of SOC. The results are reported in Fig. \ref{fig:energysoc}. 
Comparing these results with the ones without SOC, we obtain a reduction of U$_C$ to a value around 1.1 eV. For the rest, the calculation results reveal that SOC induces minor changes in the ground state energies of the four magnetic systems, although the overall trend remains largely unchanged. FM and AFM2 remain the most stable configurations. Similar to the case without SOC, the energy difference between the AFM1 and NM states is nearly identical at U=0, even when SOC is included, likely due to the small magnetic moment (0.27 $\mu_B$ with SOC). This will lead to a small $J_{\parallel}$ compared to the results for other values of U. 

Employing self-consistent energy differences, we calculate the magnetic exchanges {$J_\perp$} and {$J_\parallel$} as defined in the previous sections for both relativistic and non-relativistic calculations. The structure used in the calculation was obtained from optimization under the AFM2 magnetic configuration. 
We compare the value of exchange constants $J_\parallel$ and $J_\perp$ as a function of U, with and without SOC.
Our exchange energy calculations provide a comprehensive explanation for the magnetic ground state energetics. Disregarding the small variation in U and the effects of SOC, the absolute value of $J_\parallel$ is one order of magnitude greater than that of $J_\perp$, demonstrating that the out-of-plane exchange energy is nearly significantly higher than the in-plane exchange energy. Therefore, we have a strongly anisotropic magnetism. From the perspective of SOC, it has a slight effect on the in-plane exchange energy $J_\perp$, favoring the ferromagnetic state as the ground state.
The significantly negative out-of-plane exchange interactions not only strongly favor FM and AFM2 configurations with identical out-of-plane spin orientations as the dominant ground states, but also reduce the structural stability of the out-of-plane antiparallel spin magnetic configuration. This leads to an increase in the ground state energy of both the AM and AFM1 phases and even a tendency towards the nonmagnetic state. Furthermore, the in-plane exchange interactions are negligibly small, resulting in minimal energy differences between FM and AFM2 configurations with opposite in-plane spin alignments. The influence of U is minimal on $J_\parallel$, but it significantly affects $J_\perp$, even forcing it to change sign, which further alters the magnetic ground state. For low values of U, the positive $J_\perp$ suggests that the AFM2 phase should be the ground state. However, as 
U increases, $J_{\perp}$ decreases notably, shifting the ground state from AFM2 to FM.
Therefore, as previously mentioned, it is challenging to determine whether the ground state is FM or AFM2 based on the existing calculations, as U values between 1 and 2 eV are reasonable for Ni$_{0.25}$NbSe$_2$.

\begin{table}[htbp]
\centering
\label{tab:exchange}
\begin{tabular}{|c|c|c|c|c|} \hline 
\toprule
 & 
\multicolumn{2}{|c|}{DFT+U} & 
\multicolumn{2}{c|}{DFT+U+SOC} \\ \hline
\cmidrule(lr){2-3} \cmidrule(lr){4-5} 
 {U (\si{\electronvolt})} & {$J_\parallel$ } & {$J_\perp$ } & {$J_\parallel$ 
 } & {$J_\perp$ } \\ \hline
\midrule
\rule{0pt}{15pt} 
0 & -2.79  &  0.76 & -1.99  &  0.31 \\ 
1 & -7.88  &  0.90 & -6.58  &  0.09 \\ 
2 & -9.26  &  0.03 & -7.73  & -1.10 \\ 
3 & -9.11  & -1.08 & -7.16  & -2.44 \\ \hline
\bottomrule
\end{tabular}
\caption{Values of exchange constants J$_\parallel$ and J$_\perp$ as a function of U with and without SOC. The unit of the exchange constants is in $\si{m\electronvolt}$.}
\end{table}

\begin{figure}
    \centering
    \includegraphics[width=1\linewidth]{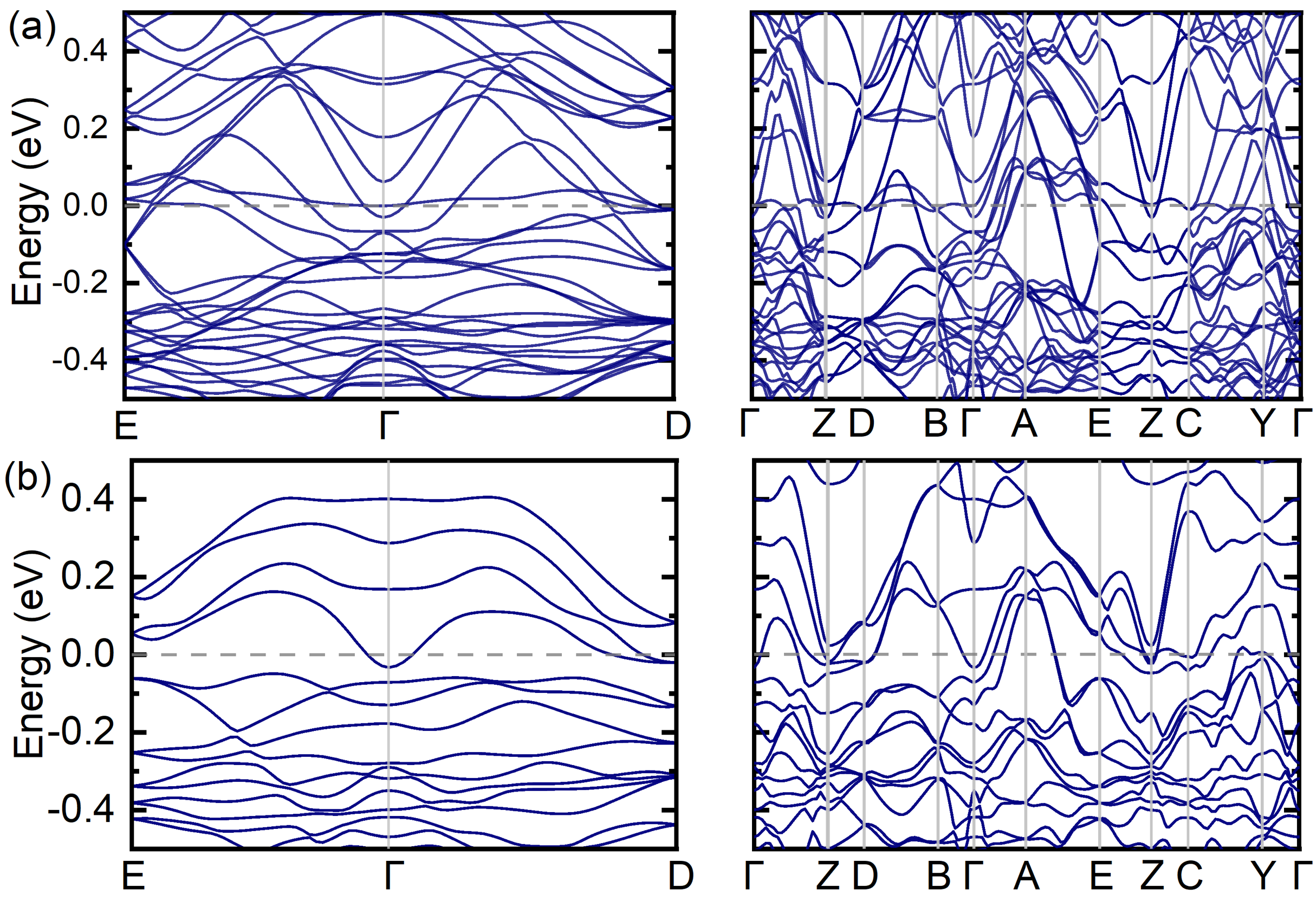}
    \caption{Electronic band structure of Ni$_{0.25}$NbSe$_2$ calculated within GGA+SOC+U for (a) the FM and (b) the AFM2 magnetic phases at U=2 eV. On the left panel is the magnification close to the $\Gamma$ point, while the right panel is the band structure across the entire k-path.}
    \label{fig:band_structure_soc}
\end{figure} 

\begin{figure}
    \centering
    \includegraphics[width=1\linewidth]{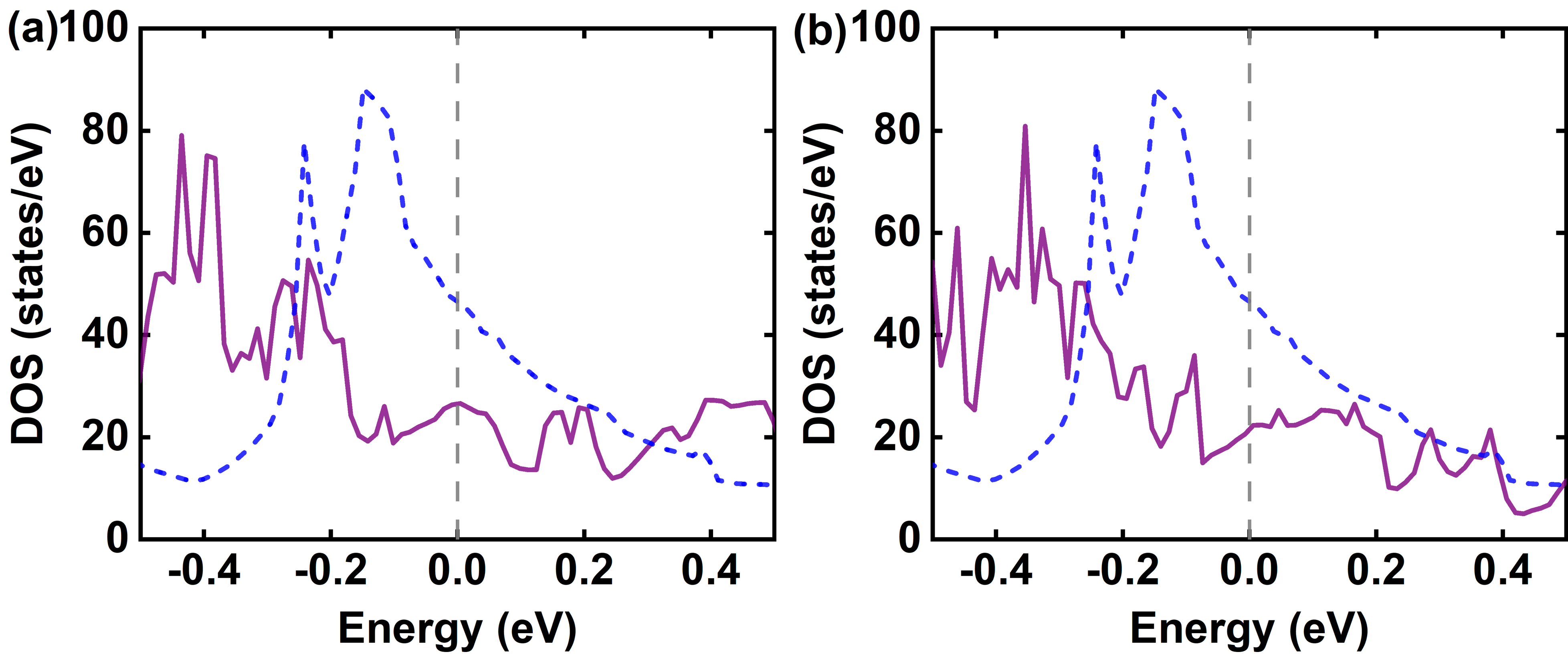}
    \caption{Total density of state (TDOS) of Ni$_{0.25}$NbSe$_2$ calculated within GGA+SOC+U for (a) the FM and (b) the AFM2 phases at U=2 eV (purple solid line). To appreciate the shift of the VHS, we included the nonmagnetic TDOS for NbSe\textsubscript{2} (blue dotted line) in both panels, with the lattice structure extracted from the literature\cite{meerschaut2001crystal}. The NbSe$_2$ TDOS is scaled to account for the different number of Nb atoms in Ni$_{0.25}$NbSe$_2$ .}
    \label{fig:dos}
\end{figure} 

\subsection{Fermiology of the AFM2 and FM phase }

We plot the relativistic band structure focusing on the most relevant magnetic phases, which are the FM phase and the AFM2 phase. Their band structures for U=2 eV are plotted in Fig. \ref{fig:band_structure_soc}(a) and \ref{fig:band_structure_soc}(b), respectively. As in the case without SOC, significant metallicity is found in both cases. In both cases, there are electron bands at the $\Gamma$ point close to the Fermi level, which is the opposite of the pristine case. The AFM2 phase hosts the Kramers degeneracy.

We plot the total density of states (TDOS) for the FM and AFM2 phases in Fig. \ref{fig:dos}(a) and \ref{fig:dos}(b), respectively, since these phases are the ground states in different regions of the phase diagram. In both cases, there is a significant DOS at the Fermi level, which produces the screening of the Coulomb repulsion, lowering the electronic correlations. We compare these DOS results with the pristine NbSe$_2$, in which there is a VHS at -0.2 eV with respect to the Fermi level, particularly due to saddle points in the electronic band structure near the M point in the Brillouin zone\cite{PhysRevB.80.241108}.
The intercalated transition metal d-orbitals bands are narrower than the Nb bands and closer to the Fermi level. Once the Nb and Ni bands hybridize, the Nb bands are pushed away from the Fermi level. The intercalation of the Ni atoms shifts the VHS to -0.4 eV from the Fermi level, as we can see in both Figs. \ref{fig:dos}(a) and \ref{fig:dos}(b). As a result, this shift of the VHS reduces the TDOS at the Fermi level and can suppress the electronic instabilities, such as the CDW and superconductivity\cite{Yan_2019}. 
It was shown that there is a strong interplay between a VHS and a CDW state in the isostructural and isoelectronic 2H-TaSe$_2$\cite{PhysRevResearch.6.013088}. Therefore, we can expect that Ni intercalation would affect the CDW. Indeed, after intercalation, the change of the Fermi surface helps stabilize the lattice. On the other hand, Ni electrons are involved in the Nb-Se covalent bonds within each monolayer, which weakens the Nb-Nb interactions and prevents the clustering effect of Nb atoms responsible for the CDW distortion. Thus, because of these two factors, the CDW phase at low temperature is expected to be suppressed in the intercalated NbSe$_2$.\cite{PhysRevB.108.L041405}. 

Another property of this family is the large orbital magnetic moment. At U=2 eV for the FM and AFM2 phase, the spin moment is around 0.9 $\mu_B$. In these phases, the orbital magnetic moment is parallel to the spin-moment and it is 0.09 and 0.08 $\mu_B$ for the FM and AFM2, respectively. Therefore, the orbital magnetic moment is of the order of 10\% with respect to the spin-moment.

\subsection{Non-collinear phases due to magnetic exchanges}

\begin{figure}
    \centering
    \includegraphics[width=1\linewidth]{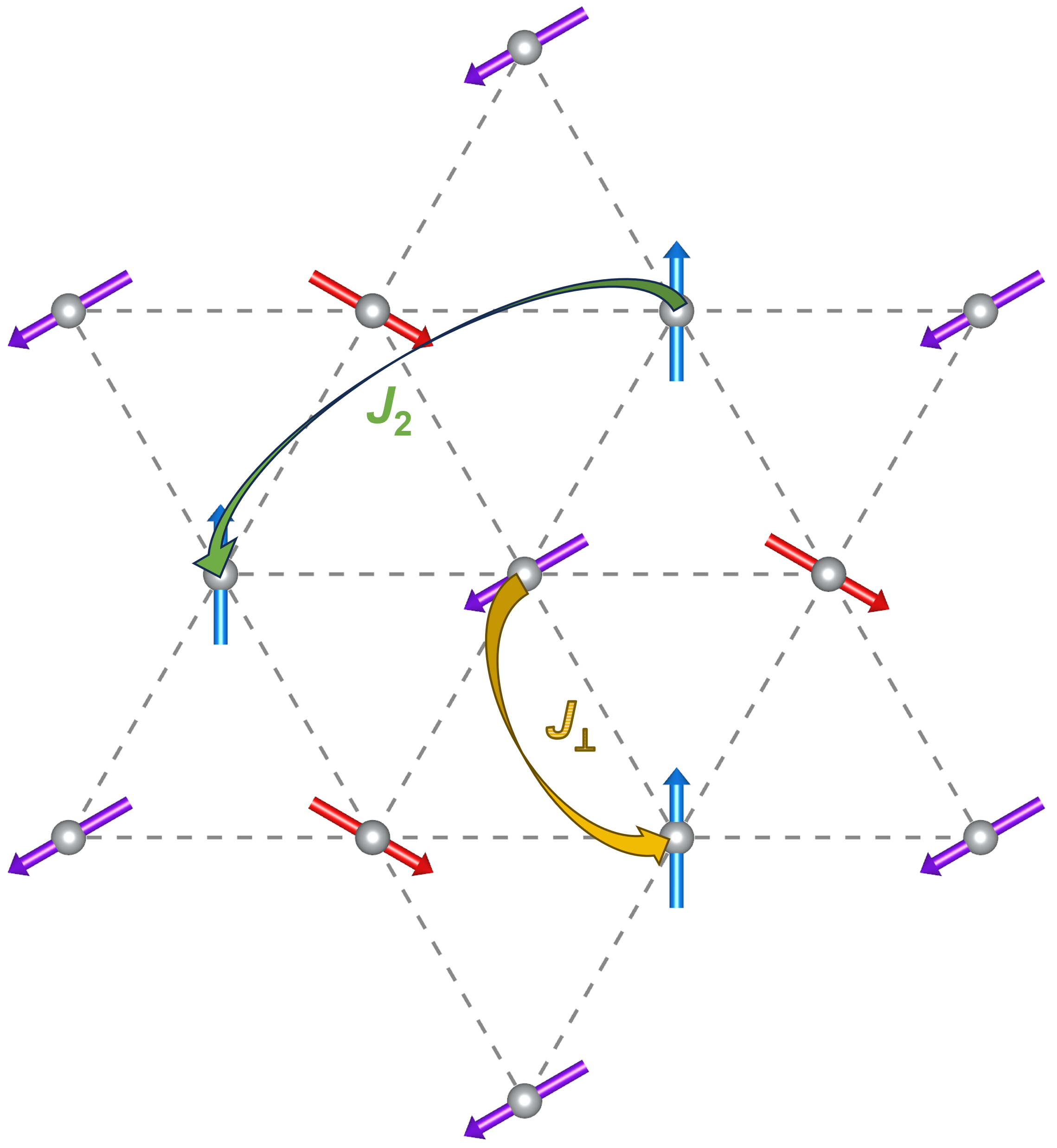}
    \caption{Non-collinear AFM-type $120^{\circ}$ phase for the triangular structure in the $ab$-plane simulated with a $\sqrt{3}\times\sqrt{3}\times1$ supercell. We plot the $ab$-plane only, as the interlayer couplings remain ferromagnetic for every value of the Coulomb repulsion. $J_{\perp}$ is antiferromagnetic. A ferromagnetic or a weak antiferromagnetic second nearest neighbour $J_2$ is necessary to stabilize this phase.}
    \label{fig:120} 
\end{figure}

The magnetic coupling is antiferromagnetic within the plane; therefore, our system is an effective 2D triangular lattice with antiferromagnetic first-neighbor interaction for low U and ferromagnetic first-neighbor interaction for high U. The collinear antiferromagnetic phase investigated in the previous section is the stripe phase. 
From the literature, we know that the ground state of a 2D triangular system is the stripe when the second neighbor magnetic exchange $J_2$ is antiferromagnetic above a threshold which is around $\frac{J_2}{J_{\parallel}}$=0.14; while the ground state is non-collinear 120$^\circ$ phase when the second neighbor interaction is ferromagnetic or weakly antiferromagnetic\cite{PhysRevB.92.140403}. Between these two phases, the theory predicts the quantum spin liquid phase for insulating systems. Since the system is metallic, it cannot host the quantum spin liquid phase, but the competition between the stripe and the non-collinear phases holds. We constructed a $\sqrt{3}\times\sqrt{3}\times1$ supercell to simulate the non-collinear 120$^\circ$ phase, which is reported in Fig. \ref{fig:120}. We performed the structural relaxation on the non-collinear phase in the presence of SOC, and we obtained that the non-collinear phase is 15 meV per formula unit higher in energy than the stripe. 
Therefore, we can assume that the value of $J_2$ is antiferromagnetic and it is the key parameter to stabilize the stripe AFM2 phase. Another factor that assisted the magnetic coupling $J_2$ in stabilizing the stripe phase is the large magnetocrystalline anisotropy. Indeed, the stripe phase can have spins along the easy axis (z-axis), while the non-collinear 120$^\circ$ phase has in-plane spins.

\subsection{Weak ferromagnetism in the altermagnetic phase of T$_{0.25}$MX$_2$}

We study the weak ferromagnetism in the altermagnetic phases of TMX$_{2}$ and T$_{0.25}$MX$_{2}$ (T = 3d, M = 4d/5d, X= S, Se, Te). What is relevant for the weak ferromagnetism calculations is only the crystal and magnetic symmetries. Since the symmetries are the same for all the T$_{0.25}$MX$_2$ systems, their properties such as weak ferromagnetism would be the same. The only quantity that is material dependent is the magnitude of the net magnetic moment;  thus, this calculation depends on the symmetries of the magnetic space group and the main results are valid for all intercalated TMDs with this stoichiometry. The weak ferromagnetism is a characteristic of systems with zero magnetization which break time-reversal symmetry, where the spin-orbit coupling generates an antisymmetric exchange interaction, leading to spin canting and a net magnetization. The simplest mechanism that produces spin canting is the staggered Dzyaloshinskii–Moriya\cite{autieri2023dzyaloshinskiimoriya}, but higher-order antisymmetric magnetic exchanges were also reported depending on the symmetries of the system\cite{839n-rckn}.

Defining the x-axis as the direction between the magnetic atoms and the y-axis as the orthogonal direction in the ab-plane, we consider the N\'eel vector \textbf{N} orientations along the x-, y-, and the z-axis. The altermagnetic phase contains two magnetic atoms that we define as Ni$_1$ and Ni$_2$, which have magnetic moments M$_1$ and M$_2$, respectively. The N\' eel vector here is defined as the difference between the magnetizations of the two magnetic sublattices \textbf{N}=\textbf{M$_1$} - \textbf{M$_2$} of an altermagnet. In the case of TMX$_{2}$, when the N\'eel vector is along the x-axis or z-axis, we have no weak ferromagnetism. When the N\'eel vector is along the y-axis, we have weak ferromagnetism along the z-axis. This is the same case as the NiAs structure compound, such as MnTe and CrSb.

In the case of T$_{0.25}$MX$_{2}$ (T = 3d, M = 4d/5d, X=S, Se, Te), although the crystal space group remains the same, the presence of more atoms reduces the symmetries of the magnetic space group. While the Te atoms in MnTe occupy the Wyckoff position 2c, in Ni$_{0.25}$NbSe$_2$, there are none in this position. With more atoms as Nb(1), Nb(2), Se(1) and Se(2), T$_{0.25}$MX$_{2}$ is expected to have lower symmetry. Therefore, we expect more cases with weak ferromagnetism. We are not aware of any prior studies addressing weak ferromagnetism in intercalated compounds with this space group. In this case, we always have weak ferromagnetism for the N\'eel vector along z and y, while we have weak ferrimagnetism for the N\'eel vector along z. Therefore, this is different from the NiAs structure, and the anomalous Hall effect (AHE) is always allowed in this class of materials.

\begin{figure*}[htb!]
\centering
\includegraphics[width=0.31\linewidth]{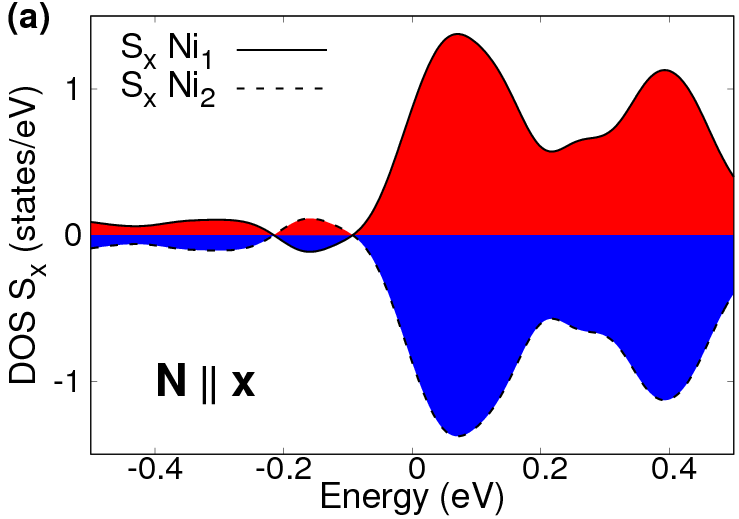} \quad
\includegraphics[width=0.31\linewidth]{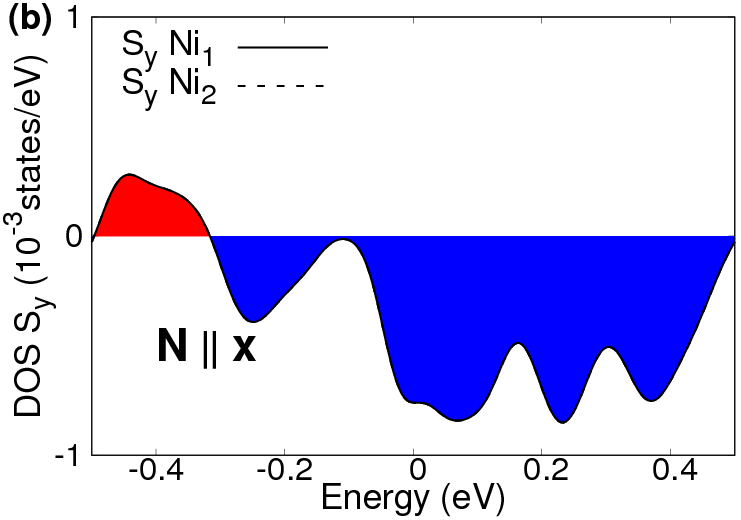} \quad
\includegraphics[width=0.31\linewidth]{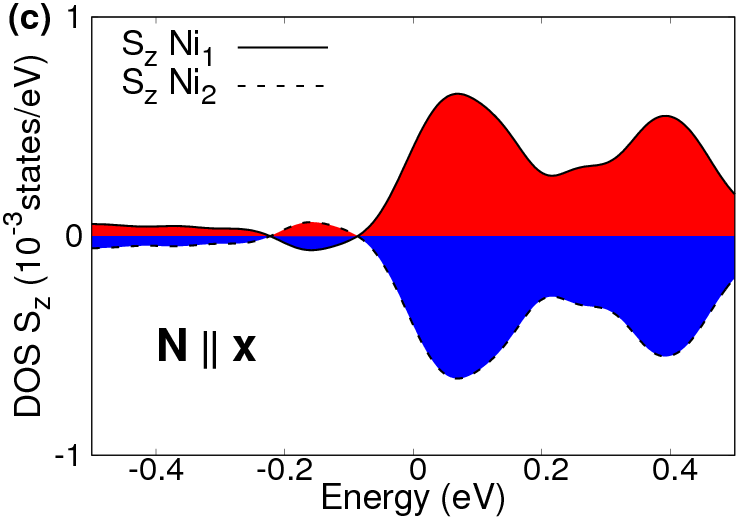}
\includegraphics[width=0.31\linewidth]{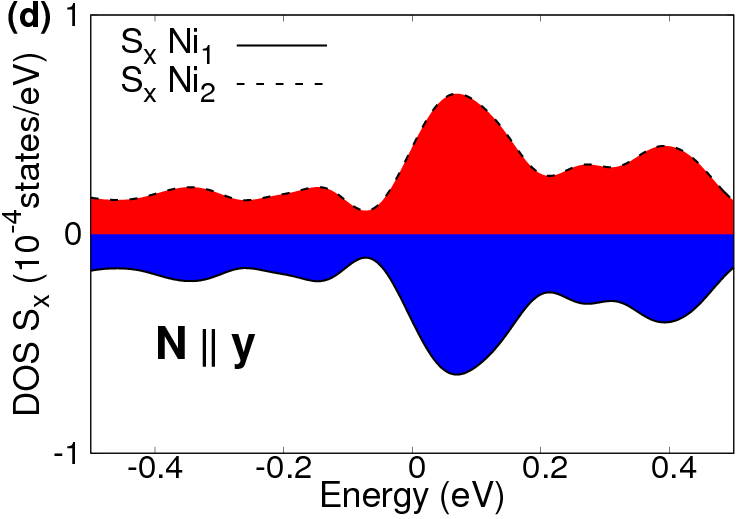} \quad
\includegraphics[width=0.31\linewidth]{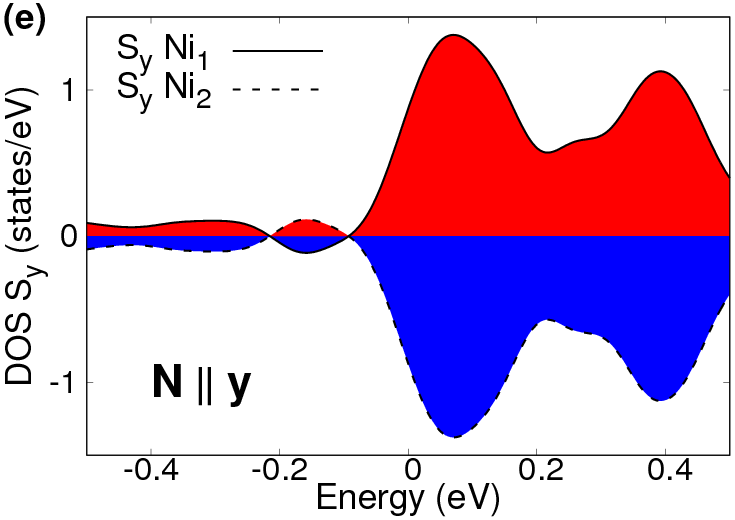} \quad
\includegraphics[width=0.31\linewidth]{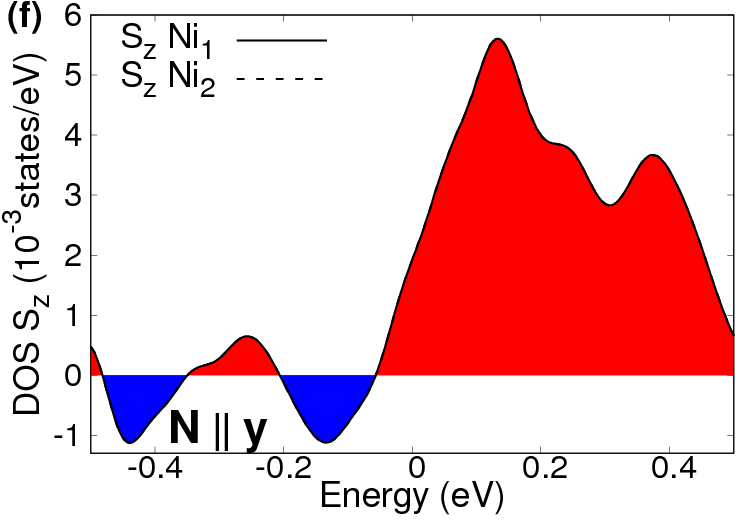}
\includegraphics[width=0.31\linewidth]{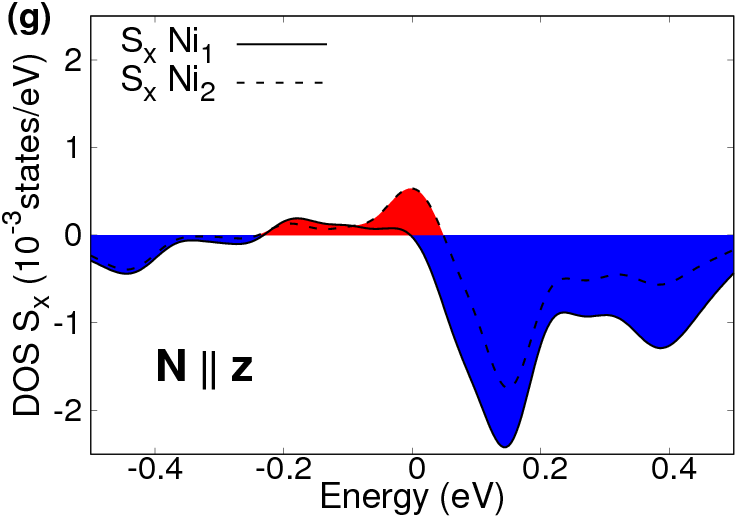} \quad
\includegraphics[width=0.31\linewidth]{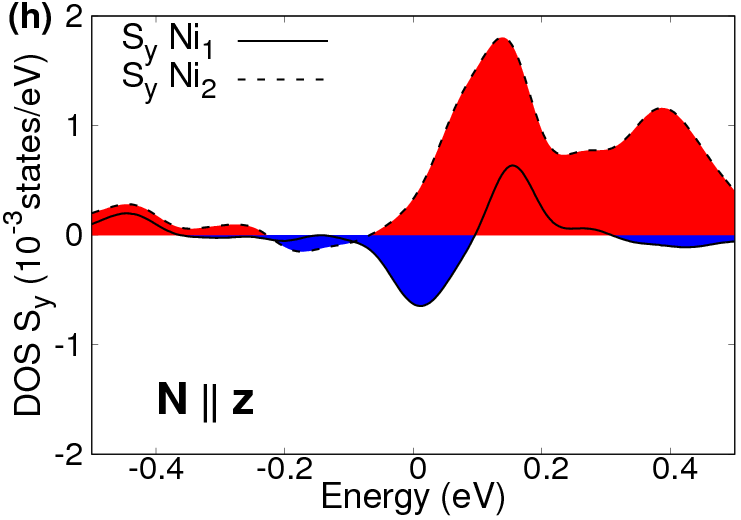} \quad
\includegraphics[width=0.31\linewidth]{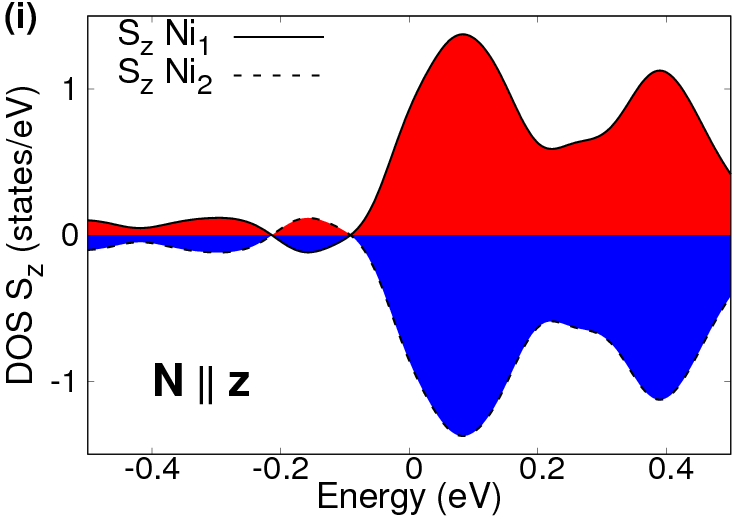}
\caption{Relativistic spin density of states between -0.5 and +0.5 eV for  Ni$_{0.25}$NbSe$_2$ for the N\'eel vector along the $x$-, $y$- and $z$-axis (\textbf{N}~$||$~\textbf{x}, \textbf{N}~$||$~\textbf{y} and \textbf{N}~$||$~\textbf{z}). (a) $S_x$ is the main component of the N\'eel vector. (b) $S_y$ component is responsible for the weak ferromagnetism along the $y$-axis. (c) $S_z$ is the component responsible for the slight rotation of the N\'eel vector in the $xz$-plane. (d) $S_x$ is the component responsible for the slight rotation of the N\'eel vector in the $xy$-plane. (e) $S_y$ is the main component of the N\'eel vector. (f) $S_z$ component is responsible for the weak ferromagnetism along the $z$-axis. (g) $S_x$ is the component responsible for weak ferrimagnetism. (h) $S_y$ is the component responsible for weak ferrimagnetism. (i) $S_z$ is the main component of the N\'eel vector.  Positive and negative spin contributions are shown in red and blue, respectively. Note that in (b) and (f) the contributions from Ni$_1$ and Ni$_2$ are identical and on top of each other.
}
\label{fig:spindensity}
\end{figure*} 

The results of the spin cantings due to the antisymmetric exchanges in altermagnets are reported in Fig. \ref{fig:spindensity}. We report the spin-density, but from the properties of the spin-density, we can infer the symmetry property of the spin-components.
When the N\'eel vector is along the x-axis, the relativistic spin-density of S$_x$ for Ni$_1$ and Ni$_2$ are equal and opposite; therefore, we have S$_x^{Ni_1}$=-S$_x^{Ni_2}$ as reported in Fig. \ref{fig:spindensity}(a). The other components are around 3 orders of magnitude lower; we have 
S$_y^{Ni_1}$=S$_y^{Ni_2}$  and S$_z^{Ni_1}$=-S$_z^{Ni_2}$ as shown in Fig. \ref{fig:spindensity}(b) and (c), respectively. Therefore, we have weak ferromagnetism along the y-axis and a small rotation of the N\'eel vector along the z-axis.
When the N\'eel vector is along the y-axis, the relativistic spin density of  S$_y$ for Ni$_1$ and Ni$_2$ are equal and opposite, so we have S$_y^{Ni_1}$=-S$_y^{Ni_2}$ as demonstrated in Fig. \ref{fig:spindensity}(e). The other two spin components S$_x$ and S$_z$ are 4 and 3 orders of magnitude smaller than S$_y$ respectively, where we have S$_x^{Ni_1}$=-S$_x^{Ni_2}$  and S$_z^{Ni_1}$=S$_z^{Ni_2}$ as shown in Fig. \ref{fig:spindensity}(d) and (f). We can deduce that we have weak ferromagnetism along the z-axis and a small rotation of the 
N\'eel vector along the y-axis. 
Finally, placing the N\'eel vector along the z-axis, the relativistic spin density of S$_z$ for Ni$_1$ and Ni$_2$ are equal and opposite as can be seen in Fig. \ref{fig:spindensity}(i). Similarly, the other two components are three orders of magnitude smaller than S$_z$. This case is, however, different than the other two cases, where the other two components are neither equal nor opposite. In this case, we have weak ferrimagnetism present along the x-axis and the z-axis. 

\begin{figure}
    \centering
    \includegraphics[width=1\linewidth]{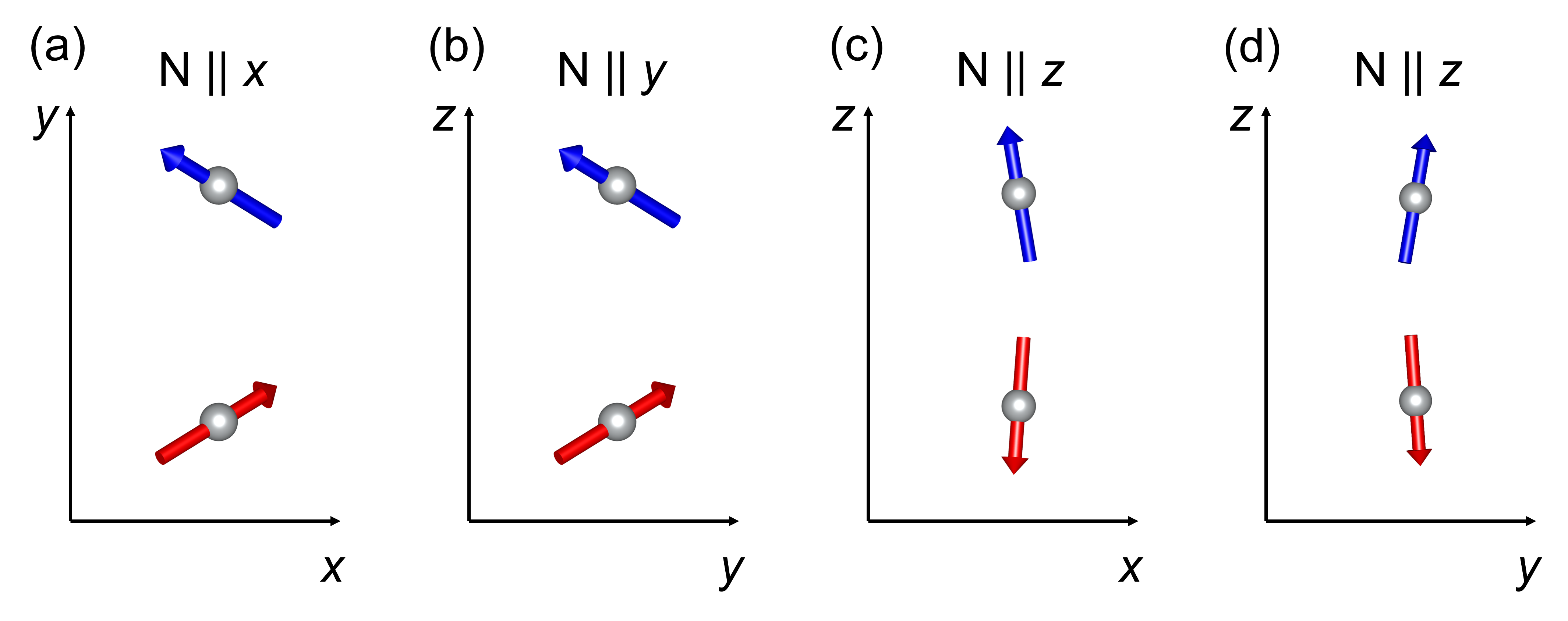}
    \caption{Summary of the spin cantings in T$_{0.25}$MX$_{2}$ (T = 3d, M = 4d/5d, X= S, Se, Te). (a) With N\'eel vector along the x-axis, we have weak ferromagnetism along the y-axis. (b) With N\'eel vector along the y-axis, we have weak ferromagnetism along the z-axis. (c,d) With N\'eel vector along the z-axis, we have weak ferrimagnetism with inequivalent spin-components along the x- and y-axis. The x-axis is defined as the Ni-Ni in-plane direction.}
    \label{summary_wFM}
\end{figure}

The summary of the weak ferromagnetism and the weak ferrimagnetism is reported in Fig. \ref{summary_wFM}. In all cases, the main components are equal and opposite; therefore, the weak ferromagnetism in this material class is observed only orthogonal to the N\'eel vector and the spins are three orders of magnitude smaller than the main components. Regarding the AHE (which has the same symmetries as the weak ferromagnetism), we can observe from Fig. \ref{summary_wFM}(a,b) that $\sigma_{yz}\ne$0 for N$\parallel$x, while  $\sigma_{yz}\ne$0 for N$\parallel$y. In the last case with N$\parallel$z, we have $\sigma_{yz}\ne$0 and $\sigma_{xz}\ne$0 as shown in Fig. \ref{summary_wFM}(c,d). In the previous subsection, the noncollinearity originated from the frustration of the symmetric magnetic exchanges, while the phases studied in this subsection are non-collinear due to the antisymmetric exchange interactions.

While in the literature orbital ferromagnetism in altermagnets was highlighted\cite{ye2025dominantorbitalmagnetizationprototypical,PhysRevLett.134.196703}, this class of altermagnets shows orbital antiferromagnetism. Tuning the N\'eel vector, we obtain three inequivalent cases with large orbital moments in directions orthogonal to the spins when the N\'eel vector is in the ab-plane and parallel to the spins when the N\'eel vector is along the z-axis. When the N\'eel vector is along the x-axis, we have weak ferromagnetism along the y-axis with the spin magnetic moment of the order of 0.001 $\mu_B$; however, the orbital moment is 0.07 $\mu_B$ along the y-axis.
For the N\'eel vector along the y-axis, we have weak ferromagnetism along the z-axis; however, the orbital moments are along the x-axis with the size of 0.07 $\mu_B$. Finally, when the N\'eel vector is along the z-axis, the orbital magnetic moment is 0.1 $\mu_B$ parallel to the spin moment. As found in the literature for other altermagnets, the orbital magnetic moment is 1-2 orders of magnitude larger than the weak ferromagnetism\cite{ye2025dominantorbitalmagnetizationprototypical,PhysRevLett.134.196703}. Also in this case, the orbital moment is around 10\% of the total spin moment.

\section{Conclusions}

We performed first-principle calculations for the Ni-intercalated NbSe$_2$ focusing on the stoichiometry Ni$_{0.25}$NbSe$_2$. We observe a strong stability of the intercalated Ni atoms to be in the middle between two Nb atoms with different z-coordinate but the same x- and y-coordinates. Our results indicate that the magnetic properties of this system are highly tunable and capable of hosting a rich variety of magnetic phases. 

Ni$_{0.25}$NbSe$_2$ is at the border between a ferromagnetic state and a stripe-like antiferromagnetic phase with Kramers' degeneracy (referred to as the AFM2 phase). This configuration is characterized by strong out-of-plane ferromagnetic coupling and relatively weak in-plane magnetic interactions. Among the magnetic structures considered, the AFM2 phase emerges as the ground state in the regime of low Coulomb repulsion, consistent with experimental conditions, where screening effects and a deviation from half-filling reduces the effective interaction strength.
The system exhibits an easy axis along $z$ and a strong in-plane second-nearest-neighbor antiferromagnetic interaction. 
We calculate the electric properties of Ni$_x$NbSe$_2$ (x=0.25, $\frac{1}{3}$ ,and 1) and found that the Ni connectivity strongly influences the ground state. When the Ni atoms belonging to two different layers are one on top of each other, the magnetic coupling between the layers is ferromagnetic. When the Ni atoms belonging to two different layers are shifted, the magnetic coupling between the layers is antiferromagnetic and the ground state can be altermagnetic.
Ni intercalation significantly alters the electronic structure near the Fermi level and makes the electronic properties more three-dimensional with respect to the pristine NbSe$_2$. It shifts the VHS away from the Fermi energy, reducing the TDOS and suppressing electronic instabilities. Nevertheless, the system remains metallic, with a substantial DOS at the Fermi level that provides effective screening of Coulomb interactions. Notable changes are observed in the Fermi surface compared to pristine NbSe$_2$, with multiple sheets appearing in both the FM and AFM2 phases. In the AFM2 and FM phases, there is a switch in the type of charge carriers at the $\Gamma$ point where an electron pocket appears at $\Gamma$ dominated by 3z$^2$-r$^2$ orbital character, with the primary atomic weight localized on the Nb atoms.

Finally, we report the study of the weak ferromagnetism for Ni$_{0.25}$NbSe$_2$. This can serve as benchmark for  the family members of T$_{0.25}$MX$_{2}$ (T = 3d, M = 4d/5d, X= S, Se, Te). 
We find weak ferromagnetism for an in-plane N\'eel vector, while weak ferrimagnetism appears for the out-of-plane N\'eel vector.
This differs from the NiAs structure, which has the same space group but not the same magnetic space group. We conclude that the family of intercalated T$_{0.25}$MX$_{2}$ hosts AHE for every orientation of the N\'eel vector. Additionally, we find a large orbital magnetic moment with antiferromagnetic order, which differs from previously studied altermagnets where orbital ferromagnetism was found.

We consider what may occur as we move to more complex cases. In the case of disordered Ni distribution or deviation of the Ni content from 1/3 and 1/4, the fact that Ni intercalated NbS$_2$ has an antiferromagnetic coupling on the triangular lattice in the ab-plane and out-of-plane interactions likely drives experimental samples into a spin-glass regime or magnetically disordered system. Given the geometrical frustration and proximity to both ferromagnetic and antiferromagnetic regimes, a metamagnetic transition under an applied magnetic field is also plausible. Assuming a proportionality between the Ni concentration and N\'eel temperature, the critical temperature is expected to be below 84 K, which is the N\'eel temperature of Ni$_{1/3}$NbS$_2$\cite{PhysRevB.108.054418}. We look forward to this magnetic system's experimental realization, which offers a promising platform for exploring tunable magnetism and correlated electronic behavior.

\begin{acknowledgments}
The authors acknowledge A. Wadge, L. Plucinski, A. Wi\'sniewski, L.  K. Tenzin, J. S\l{}awi\'nska, M. Cuoco, F. Mazzola, D. Romanin and M. Calandra for useful discussions. C. A. thanks R. Sattigeri for checking the sign of J$_{||}$ for Ni$_{0.25}$NbSe$_2$ within Quantum Espresso.
This research was supported by the "MagTop" project (FENG.02.01-IP.05-0028/23) carried out within the "International Research
Agendas" programme of the Foundation for Polish Science, co-financed by the
European Union under the European Funds for Smart Economy 2021-2027 (FENG). C.A. acknowledges support from PNRR MUR project PE0000023-NQSTI.
We further acknowledge access to the computing facilities of the Interdisciplinary Center of Modeling at the University of Warsaw, Grant g91-1418, g91-1419, g96-1808 and g96-1809 for the availability of high-performance computing resources and support.  We acknowledge the access to the computing facilities of the Poznan Supercomputing and Networking Center, Grants No. pl0267-01, pl0365-01 and pl0471-01.
\end{acknowledgments}

\medskip

\appendix

\section{Computational details}
We conducted calculations using the Vienna $ab$ $initio$ Simulation Package (VASP)\cite{kresse1993ab,kresse1996efficiency} based on density functional theory (DFT) by the projector-augmented wave (PAW) method\cite{kresse1999ultrasoft}. The exchange-correlation functional was described using the generalized gradient approximation (GGA) proposed by Perdew, Burke, and Ernzerhof (PBE)\cite{perdew1996generalized}. 
To investigate the relationship between the ground state and the on-site Coulomb interaction U in detail, the DFT $+$ U method was involved\cite{liechtenstein1995density}, U was varied from 0 to 3 eV. Hund’s coupling J$_{H}$ was set to 0.15U, a value deemed suitable for 3$d$ transition metal compounds. We also performed tests using Local Density Approximation (LDA) or PAW with the 3p of Ni as semi-core to compare the AM and FM phase. In LDA, we found that the system is non-magnetic for U = 6 eV and below. Once the system becomes magnetic, we still obtain that the FM phase is the ground state. Including the 3p of Ni still does not alter the energy relationship between the FM and AM magnetic phases.
The k-point grid was set to $10 \times 10 \times 5$ for the primitive cell of Ni$_{0.25}$NbSe$_2$ and Ni$_{0.33}$NbSe$_2$, $12 \times 12 \times 3$ for NiNbSe$_2$, and  $5 \times 10 \times 5$ for the supercell of Ni$_{0.25}$NbSe$_2$, with a total energy convergence criterion of $10^{-4}$ eV and a cutoff energy of 450 eV. To calculate the magnetocrystalline anisotropy, we use a large k-grid $16 \times 16 \times 8$ for the unit cell. NiNb$_{4}$Se$_{8}$ crystallizes in the space group  P6$_3$/$mmc$ (no. 194). We need to double the lattice parameters in the $ab$-plane using a 2 $\times$ 2 $\times$ 1 supercell; the lattice parameters of the supercell determined experimentally\cite{wadge2025b} become $a_{sup}$=$b_{sup}$=2$a$=6.9100{\AA} and $c$=12.3238 {\AA}, this supercell is used to study the FM and AM phase. We need to double the cell further and construct the supercell 4 $\times$ 2 $\times$ 1 to study the AFM1 and AFM2 magnetic configurations; this last supercell holds the space group of P2$_1$. Lattice parameters are kept fixed while atomic positions are optimized for each magnetic structure. The orbital magnetic moments are local moments on the Ni atoms calculated from the output of the VASP code. For weak ferromagnetism part, details regarding the relativistic spin density of state are reported in the literature\cite{autieri2023dzyaloshinskiimoriya}.

\bibliography{references}
\end{document}